\documentclass{article}

\usepackage{arxiv}

\usepackage[utf8]{inputenc} % allow utf-8 input
\usepackage[T1]{fontenc}  % use 8-bit T1 fonts
\usepackage{hyperref}    % hyperlinks
\usepackage{url}      % simple URL typesetting
\usepackage{booktabs}    % professional-quality tables
\usepackage{amsfonts}    % blackboard math symbols
\usepackage{nicefrac}    % compact symbols for 1/2, etc.
\usepackage{microtype}   % microtypography
\usepackage{lipsum}
\usepackage{graphicx}
\usepackage{caption} % 加载标题配置宏包
\usepackage[authoryear,longnamesfirst]{natbib}
\usepackage{caption}
\usepackage{float}
\usepackage{bm}

\usepackage{amsmath,amsfonts,amsthm,amscd}
\usepackage{graphicx}
\usepackage{subfigure}
\usepackage{tabularx}
\usepackage{amsmath}  % 数学公式核心包
\usepackage{amssymb}  % 扩展数学符号包（含\mathscr定义
\usepackage{multirow}

\DeclareMathOperator{\dist}{\mathrm{dist}}

\renewcommand{\vec}[1]{\bm{#1}}

\newcommand{\mean}[1]{\overline{#1}}

\newcommand{\set}[1]{\left\{ #1 \right\}}

\newcommand{\norm}[1]{\left\lVert #1 \right\rVert}
\newcommand{\normp}[2]{{\left\lVert #1 \right\rVert}_{#2}}
\newcommand{\trsp}[1]{{#1}^\textsf{T}}

\newcommand{\scrd}[2]{{#1}_{\mathrm{#2}}}

\newcommand{\Cum}[2]{\mathrm{Cum}_{#2}\set{#1}} % cumlative 

\usepackage{algorithm}     
\usepackage{algorithmicx}
\usepackage{algpseudocode}
\usepackage{indentfirst}
\newcommand{\Fig}{\textbf{Figure}~}
\newcommand{\Tab}{\textbf{Table}~}

\algnewcommand\algorithmicswitch{\textbf{switch}}
\algnewcommand\algorithmiccase{\textbf{case}}
\algnewcommand\algorithmicdefault{\textbf{default}}

\algdef{SE}[SWITCH]{Switch}{EndSwitch}[1]{\algorithmicswitch\ #1\ \algorithmicdo}{\algorithmicend\ \algorithmicswitch}%
\algdef{SE}[CASE]{Case}{EndCase}[1]{\algorithmiccase\ #1}{\algorithmicend\ \algorithmiccase}%
\algdef{SE}[DEFAULT]{Default}{EndDefault}[1]{\algorithmicdefault\ #1}{\algorithmicend\ \algorithmicdefault}%

\makeatletter
\renewcommand{\ALG@name}{Algorithm}
\newenvironment{breakablealgorithm}
{% begin of the breakablealgorithm
	\begin{center}
		\refstepcounter{algorithm}% New algorithm
		\setlength{\baselineskip}{15pt} 
		\renewcommand{\caption}[2][\relax]{% Make a new \caption
			\hrule height.9pt depth0pt \kern0pt
			{\raggedright\textbf{\ALG@name~\thealgorithm} ##2\par}%
			\ifx\relax##1\relax % #1 is \relax
			\addcontentsline{loa}{algorithm}{
				\protect\numberline{\thealgorithm}##2}%
			\else % #1 is not \relax
			\addcontentsline{loa}{algorithm}{
				\protect\numberline{\thealgorithm}##1}%
			\fi
			\kern2pt\hrule\kern2pt
		}
	}{% end of the breakablealgorithm
		\kern3pt\hrule\relax% \@fs@post for \@fs@ruled 
	\end{center}
}

\title{Feature Representation and Clustering of Airport Congestion with Hurst Exponent and High Order Statistics}

\author{
 Wei Sun \\
 School of computing\\
 Guangzhou College of Applied Science and Technology\\
 Zhaoqing, China 526000 \\
 \texttt{cs.weisun@foxmail.com} \\
  \And
 Zi-Feng Yi \\
School of computing\\
Guangzhou College of Applied Science and Technology\\
Zhaoqing, China 526000 \\
\texttt{yfc260@163.com} \\
 \And
 Zhi-Qiang Feng \\
 School of Information Science and Technology\\
 	Hainan Normal University\\
 Haikou, China 571158 \\
 \texttt{z.q.feng@foxmail.com} \\
  \And
 Ji Ma \\
 Sino-European Institute of Aviation Engineering\\
 Civil Aviation University of China\\
 Tianjin, China 300300 \\
 \texttt{jma@cauc.edu.cn} \\
   \And
 Ruo-Shi Yang \\
 College of Civil Aviation\\
 Nanjing University of Aeronautics and Astronautics\\
 Nanjing, China 210016 \\
 \texttt{yangruoshi@nuaa.edu.cn} \\
}

\begin{document}
\maketitle
\begin{abstract}
Air traffic controllers benefit from referencing historical dates with similar complex air traffic conditions to identify potential management measures and their effects, which is critical for understanding air transportation system laws and optimizing decisions. This study conducted data mining using flight timetables. It first explored airport congestion mechanisms and quantified congestion as time series, then proposed a higher-order cumulants based time series feature extraction method. This method was fused with other features to build a high-dimensional airport congestion feature vector, and finally K-means clustering was applied to extract and analyze congestion patterns. The clustering method was empirically validated with 2023 flight data from Guangzhou Baiyun International Airport and it accurately classified airport operational states. To verify universality, the same framework was applied to 6 airports under the "one-city, two-airports" layout in Beijing, Shanghai and Chengdu. Results showed significant congestion pattern differences between existing and newly constructed airports. Conclusions confirm the proposed feature extraction and clustering framework is effective and universal, and it can accurately capture airport congestion dynamics. Under the "one-city, two-airports" layout, existing and newly constructed airports differ significantly in operational modes, and most single-airport city airports have operational modes highly consistent with existing airports. This study provides valuable decision-making references for airport managers and air traffic controllers. It helps them deepen understanding of air traffic dynamics and airport congestion patterns, thereby optimizing traffic management strategies and improving airport operational efficiency.
\end{abstract}

% keywords can be removed

\section{Introduce}
With the development of air transportation, the contradiction between demand and supply of air traffic has become increasingly prominent, leading to severe airport congestion and flight delays. This imposes a high cost on both airlines and passengers. The delay caused by airport congestion not only dissuades passengers from choosing air transportation or opting for the same airline in the future\cite{folkes1987field,britto2012impact}, but also compels airlines to bear additional expenses for aircraft maintenance and underutilization of their fleets\cite{VLACHOS20141}. Moreover, flight delays contribute to heightened fuel consumption and carbon dioxide emissions, thus posing environmental risks. Projections from Airbus\cite{airbus2017global}, Boeing\cite{Boeing2015}, and the International Air Transport Association indicate that passenger and air cargo volume will double in the next 20 years. As air traffic demand continues to grow strongly, the congestion problem in the entire air transport system will become more severe. Therefore, it is particularly important to delve into the relationship between flight delays and the supply and demand of airspace resources.

A large number of scholars have conducted research in the field of flight delays, forming a relatively complete theoretical system and achieving abundant research results. Flight delays arise from a combination of subjective and objective factors. Subjective factors encompass airline planning, flight scheduling, and rotation adjustment. Objective factors include weather conditions, network resources, military activities, facility support, air traffic control, airport construction, as well as hardware and software resource allocation. However, the root cause of flight delays ultimately stems from the imbalance between airspace resource capacity and demand\cite{dong2013CA}. Clockner et al. argue that the primary cause of flight delays is the combination of high passenger volume and insufficient runway capacity\cite{smith2004continuous}, and they provide recommendations to increase runway capacity. The diversity of causes of flight delays makes it extremely challenging to understand the fundamental patterns of flight delays and design appropriate strategies. Many past studies have developed algorithms and simulation research that incorporate randomness in decision-making, such as weather, capacity, and demand. However, it has been demonstrated that these methods struggle to gain acceptance from air traffic experts, who generally rely on their own judgment and experience rather than recommendations from research tools\cite{gorripaty2017identifying}. To enable air traffic management experts to fully leverage their experience, the key is to identify dates in the historical data that are similar to the current situation and consider the outcomes of actions taken on those dates. Therefore, by utilizing data mining algorithms and utilizing historical weather, traffic conditions, and capacity data to identify similar dates, valuable data support can be provided to understand flight delay patterns and design appropriate strategies. Grabbe et al.\cite{grabbe2013similar} employed the k-means clustering algorithm to identify dates with similar national airspace system operations and elucidated the primary causes of ground delay programs. Meng Huifang\cite{2016MengDelayExtract} developed a combined quantitative and qualitative method for extracting flight delay patterns using nonlinear dynamical theory. Zhang M et al.\cite{2021Characterizing} transformed a substantial amount of data into spatiotemporal tensors and employed a probabilistic decomposition framework to analyze delay characteristics in different time and space segments.

Numerous scholars have extensively examined patterns of flight delays. However, the current emphasis primarily revolves around identifying and addressing delay issues, neglecting the evaluation of overall airport operational efficiency and resource utilization based on delay patterns. Consequently, our intention is to explore the operational situation of airports during similar dates by identifying congestion patterns at airports. Currently, numerous scholars discuss the causes of airport congestion from different perspectives. Nombela et al.\cite{nombela2004internalizing} believe that airport congestion is the result of joint decision-making by airport managers and airlines. Berster et al.\cite{berster2015increasing} investigate whether increasing the number of large aircraft and seat capacity at congested airports is a reasonable approach to cope with airport congestion, and the results show that it is effective for most congested airports. By increasing the number of large aircraft and capacity, the airport can reduce the number of flights that need to be processed per hour, thereby maintaining a lower level of congestion. Simaiakis et al.\cite{simaiakis2014demonstration} improve airport operational efficiency and capacity by optimizing the thrust rate at boarding gates to prevent airport congestion. Many economists\cite{vaze2012modeling,czerny2014airport} believe that efficient allocation of airport resources and reducing traffic demand during peak hours can be achieved by setting reasonable pricing for time slots. In short, airport congestion stems from the imbalance between airport traffic demand and airport capacity, and the degree of congestion can be gauged by comparing these disparities. Current air traffic management systems, such as FAA's(Federal
Aviation Administration) ETMS (Enhanced Traffic Management System), assess the level of airport congestion by comparing thresholds of traffic demand and airport capacity. It can be observed that the common cause underlying airport congestion and flight delays is the imbalance between capacity and flow. However, flight delays mainly occur at airports, which leads to even more scarce airport resources. And the scarcity of airport resources further exacerbates flight delays. 

\begin{figure*}[htbp]
	\centering
	\includegraphics[width=\textwidth]{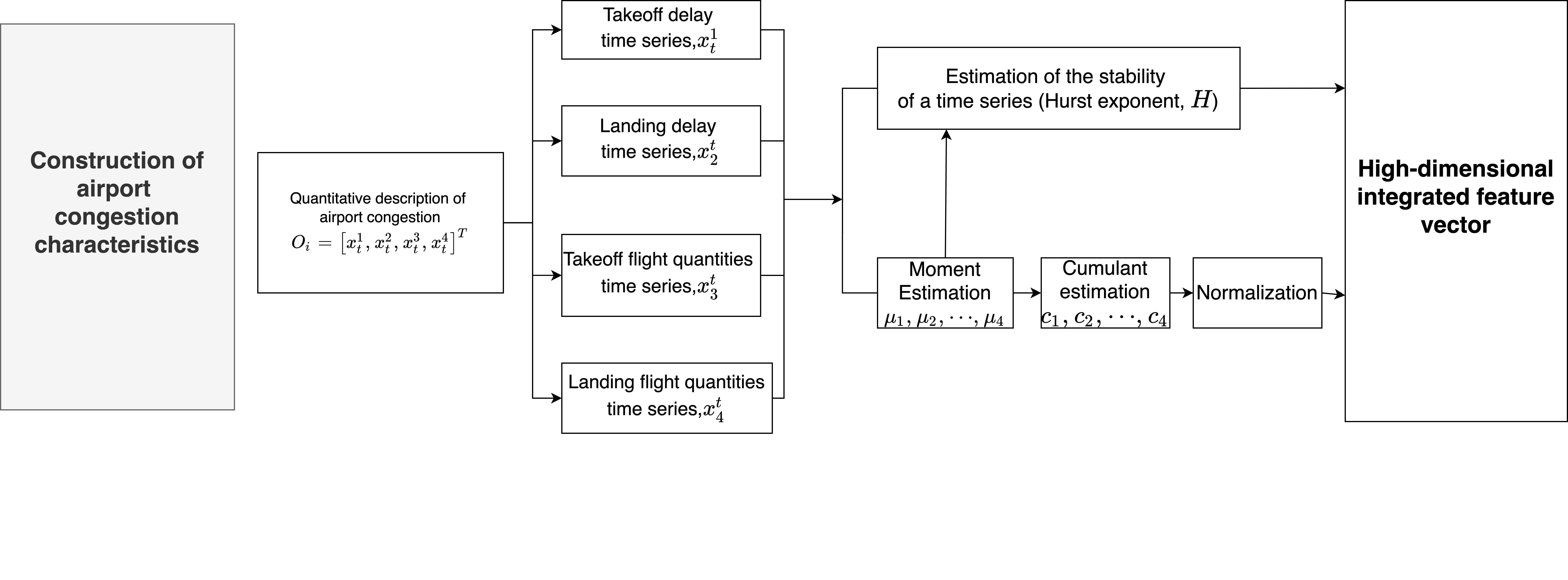}
	\caption{Technical route for extracting flight delay patterns}
	\label{Flow}
\end{figure*}

Therefore, in order to explore the congestion-delay pattern at the airport, this article takes the perspective of airport congestion and constructs a time series of flight congestion patterns. The number of aircraft takeoffs and landings and the duration of aircraft delays are used as quantitative indicators to describe the characteristics of airport congestion. To better extract the time series characteristics of congestion more effectively, we propose a time series feature extraction method based on higher-order cumulants to eliminate the influence of random noise and construct a feature space. The K-means clustering algorithm is employed to partition the feature space and identify daily airport congestion patterns. The technical process is outlined in the \Fig \ref{Flow}.

\section{Airport Congestion}\label{sec2}
\subsection{Phenomena of Airport Congestion}

When the capacity of an infrastructure fails to meet the demand, it results in traffic congestion, and the civil aviation system is no exception to this. As the civil aviation system continues to experience significant growth, the issue of inadequate transportation capacity is anticipated to intensify in the future. Nonetheless, the airport can still maintain an adequate overall capacity if the departure and arrival times of flights are evenly distributed within its operating hours\cite{roosens2008congestion}. For instance, most runways have the capability to handle 30 to 50 takeoffs and landings per hour, meaning that if the airport operates for a full 18 hours per day, each runway can potentially accommodate approximately 250,000 takeoffs and landings per year. However, practical operations and safety restrictions often mitigate the airport's capacity to a lower level. Congestion problems at major airports primarily stem from peak periods, particularly during the morning rush hour, which can trigger a series of delays and additional delays.

To alleviate airport congestion, scholars typically address the issue from two aspects: "capacity expansion" and "demand regulation". "capacity expansion" refers to expanding airport capacity and improving operational efficiency, while "demand regulation" focuses on demand management and operational improvements\cite{jacquillat2018roadmap}. These measures are interdependent and complementary. Airports usually begin by planning capacity through demand forecasting and subsequently optimize operations at a tactical level, including ground handling and air traffic flow, to maximize airport capacity and reduce operational costs. In cases where long-term demand cannot be met, strategic measures such as enhancing demand or even expanding the airport's capacity through construction may become necessary.

\begin{figure}[htbp]
	\centering
	\includegraphics[width=0.5\textwidth]{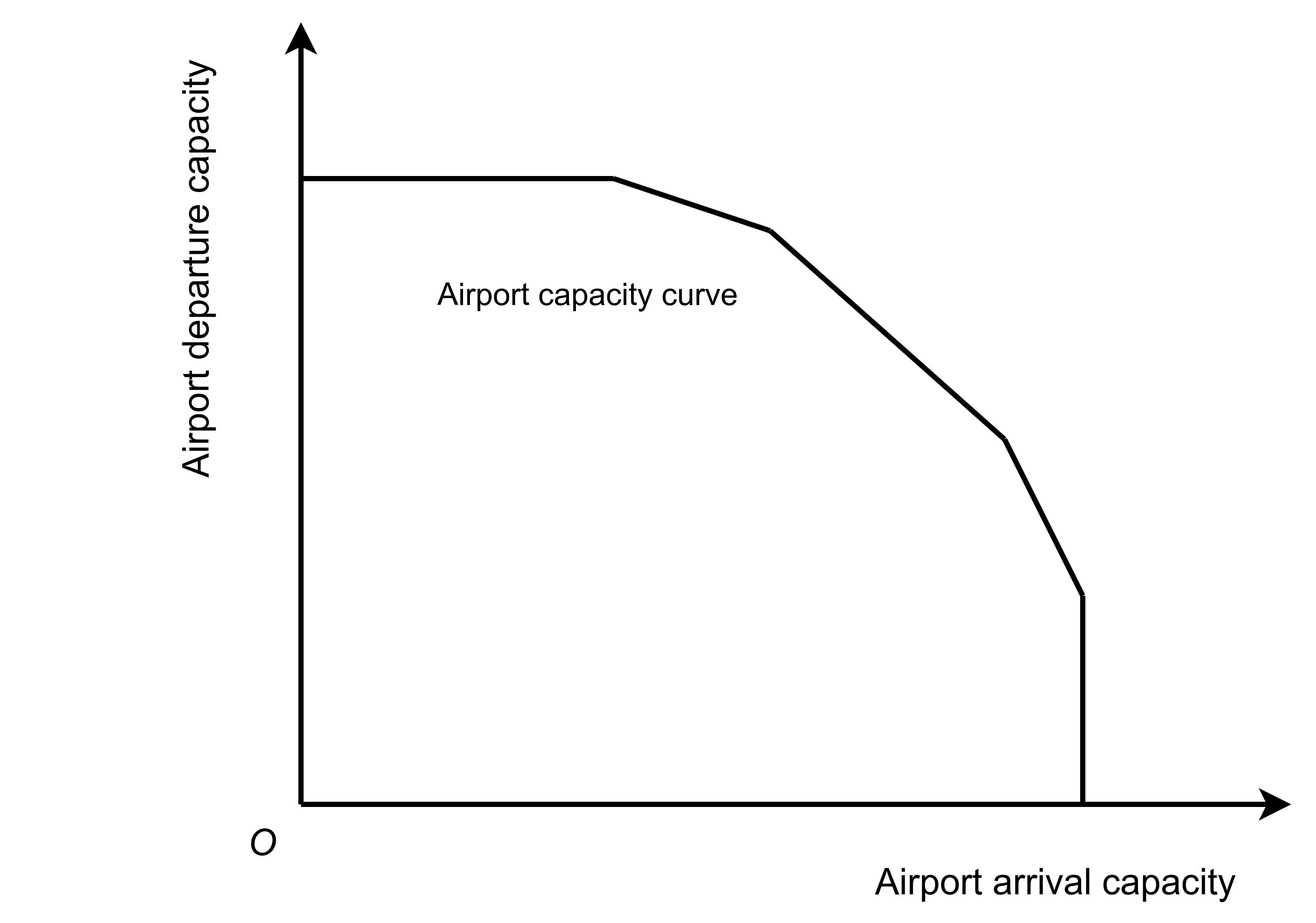}
	\caption{Capacity curve}
	\label{Capacity}
\end{figure}

\subsection{Mechanism and Parameterization of Airport Congestion}\label{sec2.2}
Airport congestion can be defined as the phenomenon of traffic congestion caused by a contradiction between transportation demand and airport capacity. When the airport capacity fails to meet the demand for arrival and departure flights, airport congestion occurs. Airport capacity is in fluenced by various factors, including weather conditions, the number of runways, and the ratio of arrival and departure flights etc. Airport capacity is divided into arrival capacity and departure capacity, often represented by the airport capacity curve shown in the \Fig \ref{Capacity} \cite{Gilbo1993airport}. When the demand for the airport is mainly composed of departure flights, the corresponding capacity is larger, and vice versa. When the airport demand exceeds capacity, congestion inevitably occurs, resulting in flight delays. Therefore, the ratio of takeoff to landing flight quantities will affect the airport capacity and consequently impact the level of congestion at the airport\cite{LiForecasting}.

\begin{figure}[htbp]
	\centering
	\includegraphics[width=0.5\textwidth]{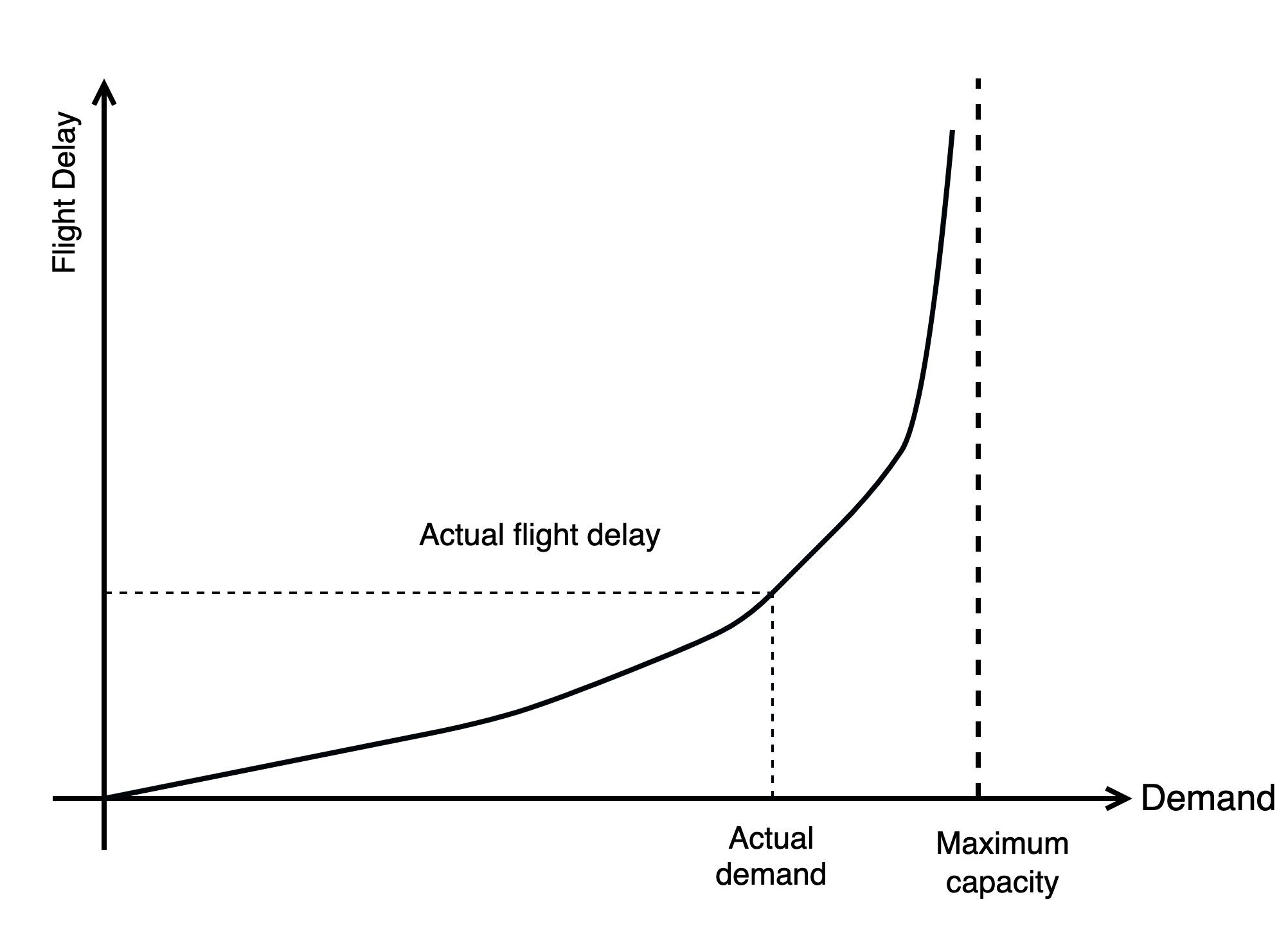}
	\caption{Variation of air transport demand with flight delays}
	\label{DeyCap}
\end{figure}

The level of congestion at the airport is closely associated with flight delays and the demand for air transport.The \Fig \ref{DeyCap} showcases the non-linear correlation between flight delays and demand. When the demand is considerably lower than the airport's maximum capacity, the rate of flight delay change is minimal, resulting in relatively low congestion at the airport. However, as the demand approaches the maximum capacity, the rate of flight delay change escalates significantly, resulting in a noticeable increase in airport congestion.

Based on the mechanistic analysis presented above, it can be concluded that airport congestion is closely associated with both the volume of flights and flight delays. Flight delays encompass both departure delays and arrival delays, which are determined by calculating the time difference between scheduled and actual departure/arrival times. A flight is categorized as delayed when its departure or arrival delay exceeds $15$ minutes. The flight delay time series of the airport can reflect the change of the airport operation capacity with time, including the number of arrival and departure flight delays, delay duration and delay fluctuation in each time scale. In actual flight operations, airlines pay particular attention to delays during specific time periods or hours of each day. In order to describe the characteristics of airport delays more accurately, we selected the time range from 05:00 to 01:00 the following day and divided it into 15-minute intervals to construct time series for departure and arrival delays time. The time series of flight quantities is constructed by the time series of departure and arrival frequencies. For convenience, we refer to the time series of delay durations for departure and arrival delays as delay time series, and the time series of takeoff and landing quantities as flight volume time series. Using $\vec{x}_t \in \mathbb{R}^{4 \times 1}$ represents the airport congestion feature vector at each moment. For the airport congestion feature time series of the $i$-th day, it is denoted as $\vec{O}_i$: 

\begin{equation}
	\vec{O}_i = \set{\vec{x}_t=\trsp{[x^1_t, x^2_t, x^3_t, x^4_t]}: 
		1\le t \le \scrd{N}{ts} }, 
\end{equation} 
for $1 \le i \le 365$ where $\scrd{N}{ts}$ represents the total number of samples in $i$-th day, $i$ represents each day within a year, and the calculation rule is: within the operational period of the airport from 
$\scrd{t}{start}=05:00$ to the time of the next day 
$\scrd{t}{end}=01:00\equiv 25:00 (\mod 24)$
which means the airport operates for $\scrd{n}{op} = \scrd{t}{end} - \scrd{t}{start} = 20$ hours each day, samples are taken every $\Delta t = 15$ minutes. Based on this, it can be inferred that
\begin{equation}
	\scrd{N}{ts} = (\scrd{t}{end} - \scrd{t}{srart})\times \frac{60}{\Delta t} = 20 \times \frac{60}{15} = 80
\end{equation}

The specific quantified indicators are presented in \Tab \ref{crowded}.
\begin{table}[H]
	\centering
	\caption{Representation of airport congestion: $\vec{x_t} = \trsp{[x_t^1, x_t^2, x_t^3, x_t^4]}$}
	\begin{tabular}{ccl}
		\toprule
		\textbf{Number} & \textbf{Symbol} & \textbf{Meaning} \\
		\midrule
		1 & $x_t^1$ & departure delay time series \\
		2 & $x_t^2$ & arrival delay time series \\
		3 & $x_t^3$ & departure flight quantities time series \\
		4 & $x_t^4$ & arrival flight quantities time series \\
		\bottomrule
	\end{tabular}
	\label{crowded}
\end{table}

\subsection{Feature Representation of Airport Congestion}\label{sec2.3}
The aim of time series extraction is to convert and condense intricate time series data into measurable and understandable representations. This conversion aids in simplifying data representation, reducing data dimensions and computational complexity, while also unveiling significant patterns and structures within the time series. For the parameterized time series of airport congestion, this study first performs feature extraction using the Hurst exponent, then further proposes a higher-order cumulants based feature extraction method to such time series. This method is designed to capture characteristics that exceed the expressive capacity of conventional statistical measures including the Hurst exponent, mean, and variance, and ultimately constructs a high-dimensional feature space specific to airport congestion.

\subsubsection{Hurst Exponent}
The Hurst exponent, proposed by British hydrologist H.E. Hurst (1900-1978), was discovered during his research on the relationship between the flow rate and storage capacity of the Nile River reservoir. He found that using biased random walks (fractal Brownian motion) could better describe the long-term storage capacity of the reservoir. Based on this finding, he proposed the rescaled range analysis method (Range/Standard deviation, R/S) to establish the Hurst exponent\cite{1951Hurst}.

For the $j$-th component sequence $\boldsymbol{O}_i^j = \{x_t^j: 1 \le t \le \scrd{N}{ts}\}$ on the $i$-th day, a sample sequence $\boldsymbol{I}_m$ with length $N_m = \left\lfloor \frac{\scrd{N}{ts}}{2^m} \right\rfloor$ (where $m = 0, 1, 2, \cdots$) is constructed. The Hurst exponent of the total sequence is then estimated using the statistical properties of the sample. For $\boldsymbol{O}_i^j$, the sample sequence $\boldsymbol{I}_m$ can be constructed as follows:

\begin{itemize}
	\item[a.] When $N_{m-1}$ is even:
	\begin{equation}
		\boldsymbol{I}_m = \left\{ \frac{x_1^j + x_2^j}{2}, \frac{x_3^j + x_4^j}{2}, \cdots, \frac{x_{N_{m-1}-1}^j + x_{N_{m-1}}^j}{2} \right\}
		\label{Eq1}
	\end{equation}
	
	\item[b.] When $N_{m-1}$ is odd:
	\begin{equation}
		\boldsymbol{I}_m = \left\{ \frac{x_1^j + x_2^j}{2}, \frac{x_3^j + x_4^j}{2}, \cdots, \frac{x_{N_{m-1}-2}^j + x_{N_{m-1}-1}^j}{2}, x_{N_{m-1}}^j \right\} 
		\label{Eq2}
	\end{equation}
\end{itemize}

In Equations \ref{Eq1} and \ref{Eq2}, $\boldsymbol{O}_i^j = \boldsymbol{I}_{m-1}$ for $m > 0$. The $\mathrm{R/S}$ statistic for the sample sequence $\boldsymbol{I}_m$ is then calculated through the following steps:

\begin{itemize}
	\item[a.] Compute the mean $E_m$ of the sample sequence:
	\begin{equation}
		E_m = \frac{1}{N_m} \sum_{i=1}^{N_m} \boldsymbol{I}_m(i) 
	\end{equation}
	
	\item[b.] Construct the deviation sequence $\boldsymbol{Y}_m$:
	\begin{equation}
		\boldsymbol{Y}_m(i) = \boldsymbol{I}_m(i) - E_m, \quad i = 1, 2, \cdots, N_m
	\end{equation}
	
	\item[c.] Calculate the cumulative deviation sequence $\boldsymbol{Z}_m$:
	\begin{equation}
		\boldsymbol{Z}_m(i) = \sum_{j=1}^{i} \boldsymbol{Y}_m(j), \quad i = 1, 2, \cdots, N_m 
	\end{equation}
	
	\item[d.] Compute the range $R_m$:
	\begin{equation}
		R_m = \max(\boldsymbol{Z}_m) - \min(\boldsymbol{Z}_m)
	\end{equation}
	
	\item[e.] Calculate the unbiased sample standard deviation $S_m$:
	\begin{equation}
		S_m = \sqrt{\frac{1}{N_m - 1} \sum_{i=1}^{N_m} \left[ \boldsymbol{I}_m(i) - E_m \right]^2}
	\end{equation}
	
	\item[f.] Compute the rescaled range $\text{RS}_m$ for this sample:
	\begin{equation}
		\text{RS}_m = \frac{R_m}{S_m}
	\end{equation}
\end{itemize}

In reference \cite{mandelbrot1969robustness}, Mandelbrot demonstrated that the $\mathrm{R/S}$ statistic asymptotically follows the relationship:
\begin{equation}
	\text{RS}_m \sim c \cdot N_m^H
\end{equation}
Where $c$ is a constant and $H$ is Hurst exponent. In logarithmic coordinates, this relationship becomes:
\begin{equation}
	\log_{10} \text{RS}_m \sim \log_{10} c + H \cdot \log_{10} N_m
\end{equation}
The Hurst exponent satisfies a linear relationship where $\log_{10} N_m$ serves as the explanatory variable and $\log_{10} \text{RS}_m$ as the response variable. After obtaining data for different samples $\boldsymbol{I}_0, \boldsymbol{I}_1, \cdots, \boldsymbol{I}_m$, the estimated value of the Hurst exponent $H$ can be derived using the least squares method.The range of values and their meanings of the Hurst exponent are as follows\cite{beran2017statistics}:
\begin{enumerate}
	\item When $0 < H < 0.5$, it means that the time series exhibits anti-persistence, or mean reversion process.
	\item When $H \approx 0.5$, it means that the time series can be described by a random walk. 
	\item When $0.5 < H < 1$, it means that the time series has memory enhancement (persistence), indicating long-term memory ability, with the strength depending on the degree of closeness to 1. As long as $H \neq 0.5$, the time series data can be described by biased Brownian motion (fractal Brownian motion) \cite{bodruzzaman1991hurst}.
\end{enumerate}

The Hurst exponent reflects the long-term correlation of delay time series and characterizes the stability of airport delays and flight volume on a given day. The vector of the Hurst exponent for the $i$-th day is denoted as $\vec{h}_i$:
\begin{equation}
	\vec{h}_i = \trsp{[h_i^1, h_i^2, h_i^3, h_i^4]}
\end{equation}
where \( h_i^j \) denotes the Hurst exponent of the \( j \)-th component sequence \( \vec{O}_i^j = \{x_t^j: 1 \leq t \leq N_{ts}\} \) on the \( i \)-th day.

\subsubsection{High-order Statistics}
Higher-order statistics generally refer to statistics that are greater than second order, including higher-order cumulants and moments. Higher-order cumulants of the third order (and above) exhibit blind Gaussianity, which means that for Gaussian sequences, their values are constantly equal to $0$ \cite{zhang2022modern}. By employing higher-order cumulant processing on non-Gaussian signals received, it is possible to automatically filter out Gaussian noise. Since the Hurst exponent has affine invariance \cite{kantz2004nonlinear}: for any constants $a$ and $b$, if the time series $\{x_t\}_{t=1}^{\scrd{N}{ts}}$ and $\{ax_t+b\}_{t=1}^{\scrd{N}{ts}}$ have a Hurst exponent, it must be the same. Therefore, sequences with the same Hurst exponent value may have different means and variances, thereby failing to showcase the non-stationary characteristics of the sequences. By incorporating the first to fourth-order cumulants of the sequence into the feature vector, the extraction of non-stationary characteristics is enhanced.

Taking $j$-th component sequence of the $i$-th day as an example, denoted as $\vec{O}_i^j = \set{\vec{x}_t^j : 1 \leq t \leq N_{ts}}$. The calculation methods for the first to fourth order cumulants are as follows\cite{Zhang1996Time}:
\begin{equation}
	\label{dup2}
	\left\{
	\begin{aligned}
		a^j_{i,1} &= \Cum{\vec{O}_i^j}{1} = \frac{1}{\scrd{N}{ts}}\sum^{\scrd{N}{ts}}_{t=1} x^j_t \\
		a^j_{i,2} &= \Cum{\vec{O}_i^j}{2} = \frac{1}{\scrd{N}{ts}}\sum^{\scrd{N}{ts}}_{t=1} \left[x^j_t - \Cum{\vec{O}_i^j}{1} \right]^2 \\
		a^j_{i,3} &= \Cum{\vec{O}_i^j}{3} = \frac{1}{\scrd{N}{ts}}\sum^{\scrd{N}{ts}}_{t=1} \left[x^j_t - \Cum{\vec{O}_i^j}{1} \right]^3\\
		a^j_{i,4} &= \Cum{\vec{O}_i^j}{4} = \frac{1}{\scrd{N}{ts}}\sum^{\scrd{N}{ts}}_{t=1} \left[x^j_t - \Cum{\vec{O}_i^j}{1} \right]^4 - 3\left[\Cum{\vec{O}_i^j}{2}\right]^2
	\end{aligned}
	\right.
\end{equation}
Where $\mathrm{Cum}_l$ denotes the $l$-th-order cumulant operator, $1\leq l \leq\scrd{k}{max} =4$, $1 \le i \le \scrd{i}{max} =365$, $1 \le j \le \scrd{j}{max}=4$.

Calculate the arithmetic mean and standard deviation of the cumulants sequence of order $1$ to order 4.
\begin{equation}
	\left\{
	\begin{aligned}
		\mu(a_{i,l}^j)&=\dfrac{1}{\scrd{i}{max}}\sum_{i=1}^{\scrd{i}{max}}a_{i,l}^j\\
		\sigma(a_{i,l}^j)&=\sqrt{\dfrac{1}{\scrd{i}{max}-1}\sum_{i=1}^{\scrd{i}{max}}\left[a_{i,l}^j-\mu(a_{i,l}^j)\right]^2}
	\end{aligned}
	\right.
\end{equation}
for $ k \in \set{1,2,3,4}$ where $\scrd{i}{max} = 365$ represents the length of the sample.

Standardize the cumulants of order $1$ to order $4$ using Z-Score normalization\cite{kreyszig2008advanced}
\begin{equation}
	\begin{aligned}
		\hat{a}_{i,l}^j=\dfrac{a_{i,l}^j-\mu(a_{i,l}^j)}{\sigma(a_{i,l}^j)},\quad 1\le l \le 4\\
	\end{aligned}
\end{equation}
and normalization
\begin{equation}
	\hat{a}_{i,l}^j=\dfrac{\hat{a}_{i,l}^j-\min\limits_{1\le i\le \scrd{i}{max}} \hat{a}_{i,l}^j}{\max\limits_{1\le i\le \scrd{i}{max}} \hat{a}_{i,l}^j-\min\limits_{1\le i\le \scrd{i}{max}} \hat{a}_{i,l}^j}, \quad 1\le l \le 4	
\end{equation}
The vector of the High order statistics for the $i$-th day is denoted as $\vec{\hat{a}}_i$:
\begin{equation}
	\vec{\hat{a}}_i = \left[ \hat{a}_{i,1}^1, \hat{a}_{i,2}^1, \hat{a}_{i,3}^1, \hat{a}_{i,4}^1, \hat{a}_{i,1}^2, \hat{a}_{i,2}^2, \hat{a}_{i,3}^2, \hat{a}_{i,4}^2, \hat{a}_{i,1}^3, \hat{a}_{i,2}^3, \hat{a}_{i,3}^3, \hat{a}_{i,4}^3, \hat{a}_{i,1}^4, \hat{a}_{i,2}^4, \hat{a}_{i,3}^4, \hat{a}_{i,4}^4 \right]^\top
\end{equation}
where $\vec{\hat{a}}_i \in \mathbb{R}^{16 \times 1}$ denotes the high order statistics of the \( j \)-th component sequence \( \vec{O}_i^j = \{x_t^j: 1 \leq t \leq N_{ts}\} \) on the \( i \)-th day.

\subsection{High-dimensional Integrated Feature Vector}
Use $\vec{f}_i \in \mathbb{R}^{20 \times 1}$ to denote the congestion feature vector of the $i$th day: 
\begin{equation}
	\vec{f}_i =\begin{bmatrix}
		\vec{h}_i\\
		\hat{\vec{a}}_i\\
	\end{bmatrix}
\end{equation}
The collection of annual congestion feature vectors forms a matrix $\mathcal{F} \in \mathbb{R}^{20 \times 365}$, where:
\begin{equation}
	\mathcal{F} = \set{\vec{f}_i: 1\le i \le \scrd{i}{max}=365}
\end{equation}

\section{Clustering Analysis of Airport Congestion}\label{sec4}

\subsection{Data Clustering: Principle and $K$-means clustering algorithm}
Cluster analysis is a fundamental method in unsupervised learning. It aims to categorize a given set of samples into distinct groups based on the similarity or distance between their features. Several widely adopted clustering algorithms are described below\cite{liStatistical}:
\begin{itemize}
	\item Distance-based clustering: the goal of this type of algorithm is to minimize intra-cluster distance and maximize inter-cluster distance. K-means clustering is a typical distance-based clustering algorithm; 
	\item Density-based clustering: this type of algorithm divides samples based on the density of their neighboring region. The most common density-based clustering algorithm is DBSCAN; 
	\item Hierarchical clustering algorithm: this includes agglomerative hierarchical clustering and divisive hierarchical clustering, among others; 
	\item Spectral clustering algorithm based on graph theory. 
\end{itemize}

In this study, we employ the K-means clustering algorithm to partition the high-dimensional integrated feature space that was constructed in Section \ref{sec2}. The dataset $\mathcal{F}$ consists of $P \times Q$ samples, where $\mathcal{F} = \{\vec{f}_i: 1 \leq i \leq \scrd{i}{max}\}$, and each $\vec{f}_i$ represents a $Q$-dimensional vector with $Q = 20$. The main objective of K-means clustering is to assign $\scrd{i}{max}$ samples to $k$ clusters. The centroid of each cluster is denoted as

% where $1 \leq k < \scrd{k}{max}=4$. 

\begin{equation}
	\vec{C}_k =\trsp{ \left[ C_{1},C_{2},\cdots,C_{k} \right]}
\end{equation}

If the Euclidean distance is utilized as the distance metric for the K-means clustering algorithm, the distance ($d_{ij}$) between samples can be defined as:
\begin{equation}
	d_{ij} = \sum_{p=1}^{\scrd{i}{max}} (f_{pi} - f_{pj})^2 = \norm{\vec{f}_i - \vec{f}_j}^2
\end{equation}

The final loss function is defined as the sum of distances between samples and their respective class centers: 
\begin{equation}
	L = \sum_{i=1}^{\scrd{i}{max}} \sum_{j=1}^{k} \delta(\vec{f}_i, k) || \vec{f}_i - \vec{C}_k ||^2
\end{equation}
Where, the function $\delta(\vec{f}_i, k)$ is an indicator function that returns a value of 1 when the sample $\vec{f}_i$ belongs to cluster $k$, and $0$ otherwise.

The K-means clustering process can be described using the pseudocode provided below.

\begin{breakablealgorithm}
	\caption{K-means}
	\begin{algorithmic}[1]
		\Require Dataset $\mathcal{F} = \set{\vec{f}_i: 1\le i \le \scrd{i}{max}}$, Number of clusters $k$
		\Ensure Representative points $C = \{C_1, C_2, ..., C_k\}$, Sample assignments $Y = \{y_1, y_2, ..., y_n\}$
		\Function{K-means}{$\mathcal{F}$,$k$}
		\State Randomly initialize $k$ representative points as initial cluster centers $C = \{c_1, c_2, ..., c_k\}$
		
		\Repeat
		\For{$i = 1$ to $i_{max}$}
		\State Compute the Euclidean distance between sample $f_i$ and each cluster center $C_k$
		\State Assign sample $f_i$ to the nearest cluster, resulting in assignment $y_i$
		\EndFor
		
		\For{$j = 1$ to $k$}
		\State Compute the mean of all samples in cluster $C_j$, resulting in new cluster center $C_t$
		\State Update cluster center $C_j = C_t$
		\EndFor
		
		\Until{convergence or satisfying stopping criteria}
		
		\State \textbf{return} Cluster center set $C$ and sample assignment set $Y$
		\EndFunction
	\end{algorithmic}
\end{breakablealgorithm}

\subsection{Assessment of Airport Congestion}\label{sec3.2}
To verify the validity of the clustering outcomes, an airport congestion evaluation index is proposed herein, which is rooted in the operational mechanism of airport congestion. Conventionally, airport capacity is defined as the maximum number of takeoffs and landings that an airport can safely and efficiently handle within a unit time under specific conditions. However, these specific conditions can hardly fully cover the complex scenarios in real-world operations. Given that the actual airport operation data used in this paper are based on the airport's actual operations under various complex conditions, it is inappropriate to simply equate the hourly planned takeoff and landing data of the airport with airport capacity. Instead, the dynamic interaction between airport capacity and traffic flow needs to be extended to the interactive influence between air traffic control capacity and air traffic flow.

Air traffic control capacity refers to the maximum number of aircraft movements that an air traffic control unit can safely handle per unit time within a specified period, within the scope of an airport or an airspace unit (e.g., a sector or air route).  This capacity is determined based on a combination of factors, including aircraft performance, constraints imposed by airport infrastructure capacity, air traffic control operational rules, acceptable delay levels, and acceptable air traffic controller workload levels. During the flight schedule adjustment in the strategic phase, the slot management department of the Civil Aviation Administration formulates the flight schedule for the next season based on airlines' slot applications and in combination with the declared capacity of the airport. Nevertheless, in the actual operational phase, due to the impact of multiple factors such as weather conditions and air traffic control restrictions, the actual operational situation often deviates from the plan. In this process, air traffic control capacity, which is evaluated during operation, comprehensively considers factors such as weather trends and other airspace users' activities, and is applied in pre-tactical and tactical traffic flow management. It serves as the core link connecting the physical capacity of the airport (e.g., runways, terminals) and actual traffic flow. It is not only constrained by the capacity of airport infrastructure but also directly determines the number of aircraft that can be safely accommodated within a specific time range.

In essence, although actual airport operational data cannot directly map to capacity and traffic flow as defined in the general sense, there is a close association between them. Moreover, actual airport operational data can more dynamically reveal the intrinsic relationship between airport capacity and traffic flow. Based on this, this study defines hourly planned takeoff and landing flights as air traffic flow adjusted by air traffic control dynamic capacity(hereinafter referred to as "dynamic traffic flow"), refers to the total number of scheduled aircraft takeoffs and landings completed at an airport per unit time, formulated by airlines based on market demand and route planning, and it represents the air traffic demand pressure faced by the airport during a specific period. The hourly actual takeoff and landing movements, defined as air traffic control dynamic capacity (hereinafter referred to as "dynamic capacity"), refers to the total number of aircraft takeoffs and landings completed by an airport per unit time under specific operational conditions. This indicator directly reflects an airport’s actual operational carrying capacity in a dynamic environment and constitutes a concrete manifestation of air traffic control dynamic capacity. To quantify the relationship between these two parameters, the following indicators are selected, with specific details presented in \Tab \ref{tab:congestion_metrics}:

\begin{itemize}
	\item $Q$: Dynamic capacity, defined as the hourly actual takeoff and landing flights (i.e., hourly air traffic flow adjusted by air traffic control capacity);
	\item $V$: Dynamic traffic flow, defined as the hourly planned takeoff and landing flights (i.e., consistent with the theoretical maximum air traffic carrying capacity of air traffic control Capacity);
	\item $\omega = \frac{|Q_i-Q_{i-1}|}{Q_{i-1}}$: Reflects the change rate of hourly actual takeoff and landing flights;
	\item $\theta = \frac{|V_i-V_{i-1}|}{V_{i-1}}$: Indicates the dynamic adjustment range of hourly planned takeoff and landing flights.
\end{itemize}
Simultaneously, from the delay perspective, two additional metrics are introduced to quantify the practical impacts of imbalances between air traffic control capacity and traffic demand: the flight punctuality rate $\gamma$ and the average delay time per flight $\tau$. The flight punctuality rate \(\gamma\) is defined as the ratio of the number of on-time flights to the total number of flights within a specific time period. Mathematically, let \(N_{\text{on-time}}\) denote the number of on-time flights and \(N_{\text{total}}\) denote the total number of flights; then \(\gamma\) can be expressed as $\gamma = \frac{N_{\text{on-time}}}{N_{\text{total}}}$. The average delay time per flight \(\tau\) is defined as the total delay time of all flights divided by the total number of flights within a given time frame. Mathematically, let \(T_{\text{total-delay}}\) denote the cumulative delay time of all flights and \(N_{\text{total}}\) denote the total number of flights; then \(\tau\) can be expressed as $\tau = \frac{T_{\text{total-delay}}}{N_{\text{total}}}$.

Through the above indicators, the operational status of the airport on a given day can be comprehensively assessed by calculating the daily average of each hourly indicator. Specifically, this assessment covers two dimensions: one is the imbalance relationship between capacity and flow, which is reflected by the daily averages of the comparison between Q and V as well as the difference in changes between \(\omega\) and \(\theta\); the other is the actual performance of flight delays, which is characterized by the daily averages of \(\gamma\) and \(\tau\). This approach provides a standardized quantitative basis for subsequent pattern recognition and clustering analysis.

\begin{table}[H]
	\caption{Airport Congestion Evaluation Metrics}
	\begin{tabular}{cclc}
		\toprule
		\textbf{Number} & \textbf{Symbol} & \textbf{Description} & \textbf{Unit}\\
		\midrule
		1 & $Q$ & Dynamic capacity, defined as the hourly actual takeoff and landing flights & Flights/hour\\
		2 & $V$ & Dynamic traffic flow, defined as the hourly planned takeoff and landing flights & Flights/hour\\
		3 & $\omega$ & Hourly change rate of traffic demand, calculated as $\omega = \frac{|Q_i-Q_{i-1}|}{Q_{i-1}}$ & Percentage\\
		4 & $\theta$ & Hourly change rate of traffic control capacity, calculated as $\frac{|V_i-V_{i-1}|}{V_{i-1}}$ & Percentage\\
		5 & $\gamma$ & Flight punctuality rate, calculated as $\gamma = \frac{N_{\text{on-time}}}{N_{\text{total}}}$ & Percentage\\
		6 & $\tau$ & Average delay time per flight, calculated as $\tau = \frac{T_{\text{total-delay}}}{N_{\text{total}}}$ & Minutes\\
		\bottomrule
	\end{tabular}
	\label{tab:congestion_metrics}
\end{table}

\subsection{Clustering Analysis of Empirical Data of Airport Congestion}\label{sec4.2}
Our research utilized data on all domestic flights in China for the year 2023, obtained from VariFlight, the~leading aviation data service provider in China. These data included the most extensive information regarding the number of flights and airports in China during that year. The~data consist of 11 valid attributes that are presented in \Tab \ref{table1}. The~redundancy of some redundant variables, which were not related to our research objectives, could not be used in our research and could potentially affect data processing speed. To~address this issue, the~researchers removed airport location information. 

\begin{table}[H]
	\caption{Description of Data.}
	\centering
	\begin{tabular}{ll}
		\toprule
		\textbf{Variable Type}     & \textbf{Variable Name}                                   \\
		\midrule
		\multirow{2}{*}{Time Information} & Departure time information: Scheduled departure time, Actual departure time    \\
		\cmidrule{2-2}
		& Arrival time information: Scheduled arrival time, Actual arrival time   \\
		\midrule
		Delay Information     & Departure delay in seconds                           \\
		\midrule
		\multirow{3}{*}{Airport Information} & Landing airport information: Airport IATA code                       \\
		\cmidrule{2-2}
		& Take-off airport information: Airport IATA code                      \\
		\cmidrule{2-2}
		& Airport location information: Airport Longitude, Airport Latitude, Province, City \\
		\bottomrule
	\end{tabular}
\label{table1}
\end{table}

In the data lifecycle , the occurrence of missing values, outliers, and inconsistent data is inevitable. To ensure the accuracy of data quality and the effectiveness of analysis, it is necessary to systematically verify missing values, outliers, and duplicate records. The dataset contains a total of 4,961,074 records, corresponding to the total number of flights throughout 2023. Among them, 35,715 records (accounting for 0.07\%) have missing actual takeoff and landing times. The study uses the direct deletion method to handle such missing values. In addition, as described in Section \ref{sec2.2}, the dataset must meet specific constraints: the actual flight departure time must be limited to the interval from 05:00 to 01:00 the next day, and only records that meet this time range are defined as valid data.

After data preprocessing, feature extraction is performed on the time series according to the method described in Section \ref{sec2.3}, and the obtained feature vectors are standardized to ensure the effectiveness of distance measurement in cluster analysis. For the high-dimensional comprehensive feature vectors integrating Hurst exponents and normalized higher-order cumulants, the Z-Score normalization method is used for transformation, so that each feature meets the distribution characteristics of mean 0 and standard deviation 1, thereby eliminating the interference caused by differences in dimensions and magnitudes between features. This standardization step can ensure the balanced contribution weight of each feature in the calculation of clustering distance and avoid systematic deviations of the results caused by dominant features.

In the clustering stage, the K-means algorithm described in Section \ref{sec4} is adopted, and the number of clusters is determined to be 4 based on practical experience. To verify the rationality of the clustering results, the study conducts analysis from both statistical and practical perspectives: at the statistical level, quantitative evaluation is carried out through intra-cluster compactness (average distance between samples in the cluster) and inter-cluster separation (distance between cluster centers). When the intra-cluster distance is small and the inter-cluster distance is large, it indicates that the clustering results are effective; at the practical level, logical inference is made on the clustering results with the help of the airport congestion evaluation indicators proposed in Section \ref{sec3.2} to verify its rationality .

This comprehensive cluster analysis framework integrating data preprocessing, feature standardization, algorithm selection, and result verification provides a systematic methodological support for mining airport congestion patterns from high-dimensional time series data.

\section{Validation and Verification}
\subsection{Beijing Capital International Airport Clustering Example}
This article analyzes the flight schedule data provided by VariFlight, which has been processed through Section \ref{sec4.2}, from January 1, 2023 to December 30, 2023, to study the flight situation at Guangzhou Baiyun International Airport(IATA Code: CAN).

According to the data, the first step is to construct a time series of crowded characteristics at the airport. To accurately depict the operational status of the Guangzhou Baiyun International Airport, all takeoff and landing flights at the airport are selected. Additionally, in line with specific application requirements, data corresponding to the time window between 01:01 and 04:59 in the early morning is excluded. Each flight is assigned to its corresponding day based on the actual departure time, and time series for delay and flight volumes are constructed. The time series for the i-th day is
\begin{equation}
	O_i = \set{\vec{x}_t=\trsp{[x^1_t, x^2_t, x^3_t, x^4_t]}: 1\le t \le \scrd{N}{ts} }
\end{equation}
The notation $\scrd{N}{ts} = 80$ represents the total number of samples taken from 05:00 to 01:00 the following day at the airport. According to Section \ref{sec2}, we extract features from each time series, resulting in obtaining the set of annual crowding feature vectors $\mathcal{F}$ with a dimension of $365\times 20$. Subsequently, the K-means clustering method is utilized to cluster the feature vectors for each day, with a parameter $k=4$ chosen based on empirical observations. The resulting clustered feature points are categorized into four distinct sets, specifically designated as $C_1,C_2,C_3,C_4$.

\subsection{Statistical Inference Evaluation of Cluster Quality}
The formula for calculating the centroid of each cluster can be expressed as:
\begin{equation}
	\mu(C_k) = \frac{1}{N_k}\sum_{\vec{f}_i \in C_k} \vec{f}_i 
\end{equation}
Where $N_k = 20$ represents the number of features in the set of congested characteristic vectors for the year.

\begin{figure}[htbp]
	\centering
	\includegraphics[scale=0.3]{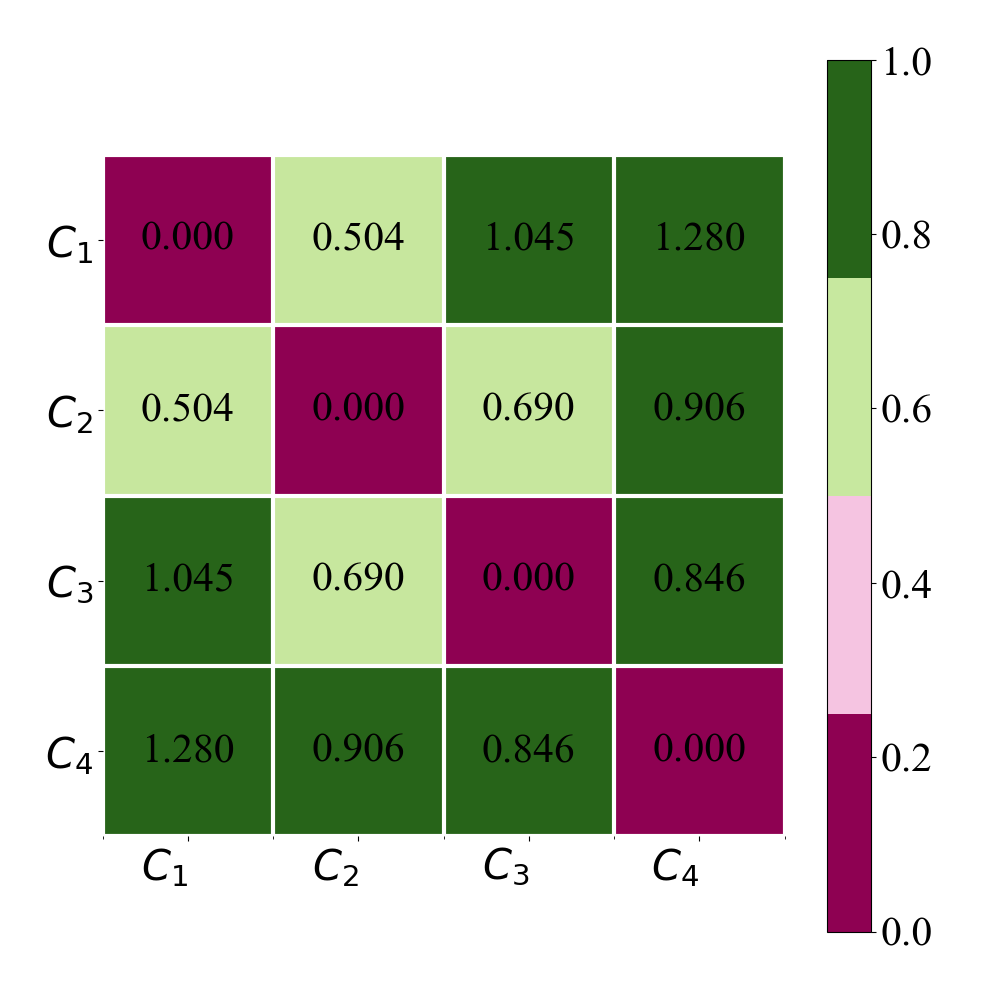}
	\caption{Euclidean Distance Matrix of Guangzhou Baiyun International Airport Clusters}
	\label{hotheat}
\end{figure}

\begin{figure}[htbp]
	\centering
	\includegraphics[width=0.5\textwidth]{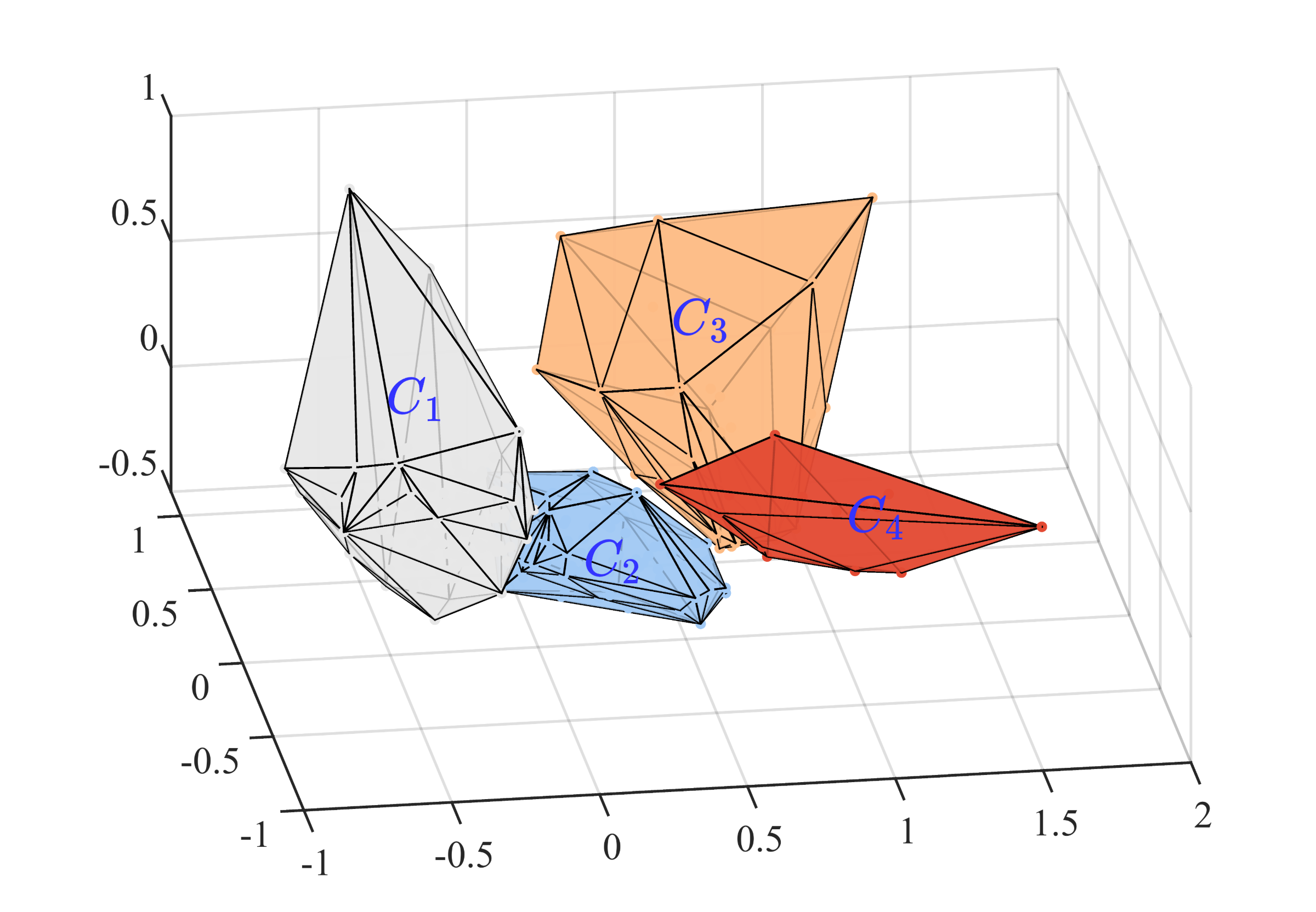}
	\caption{Visualization of Clustering Results for Guangzhou Baiyun International Airport }
	
	\label{cluster_result}
\end{figure}

The dissimilarity between two feature vectors $\vec{f}_i$ and $\vec{f}_j$ in the feature vector space can be quantified by measuring their Euclidean distance: 
\begin{equation}
	\dist_E(\vec{f}_i,\vec{f}_j) = \normp{\vec{f}_i-\vec{f}_j}{2}
\end{equation}
For clustering clusters $C_i$ and $C_j$, the dissimilarity between them is defined by measuring the degree of separation of different clusters based on the Euclidean distance of their respective centroids.
\begin{equation}
	\dist_E(C_i,C_j) = \normp{\mu(C_i) - \mu(C_j)}{2} \label{E_cluster}
\end{equation}

As shown in \Fig \ref{hotheat}, the analysis results reveal that the Euclidean distances among the four clusters are generally large. However, the Euclidean distances between cluster $C_1$ and cluster $C_2$, as well as between cluster $C_2$ and cluster $C_3$, are relatively smaller, yet they still fall within an acceptable range. To further verify the clustering effect of these four clusters, we employed Principal Component Analysis (PCA) to reduce the dimensionality of the high-dimensional feature vectors and visualized the clustering results, with the outcomes presented in \Fig \ref{cluster_result}. The visualization results indicate that even though the Euclidean distances between some clusters are relatively small, all clusters can still be effectively distinguished.

\begin{table}[b] 
	\caption{Clustering Evaluation Results for Guangzhou Baiyun International Airport}
	\centering
	\begin{tabular}{llllllll} 
		\toprule 
		$C_k$ & $\overline{Q}$ & $\overline{V}$ & $\overline{\omega}$ & $\overline{\theta}$ & $\overline{\gamma}$ & $\overline{\tau}$ & Days \\
		\midrule 
		$C_1$ & 61.8139    & 61.4440    & 0.1386        & 0.0851        & 0.7502        & 7.9863       & 181 \\
		$C_2$ & 57.4059    & 57.6513    & 0.1380        & 0.1068        & 0.7050        & 12.1177      & 136 \\
		$C_3$ & 50.0734    & 52.7624    & 0.2778        & 0.1360        & 0.5985        & 36.0798      & 33  \\
		$C_4$ & 40.2491    & 39.9474    & 0.1780        & 0.1584        & 0.8645        & 4.9089       & 15  \\
		\bottomrule 
	\end{tabular}
\label{result} 
\end{table}

%Aa3#
\subsection{Cluster Results and Interpretation}
In accordance with Section \ref{sec3.2}, we calculate the congestion evaluation indicators $Q_i, V_i, \omega_i, \theta_i$ , $\gamma_i$ and $\tau_i$for each day of the year, with $i = 1, 2, \cdots, 365$ denoting the $i$th day. Based on the clustering results, the mean value of the congestion indicators for the corresponding days is determined.

\begin{equation}
	\left\{
	\begin{aligned}
		\overline{Q} & = \frac{1}{N_{C_k}}\sum_{i \in C_k}Q_i\\
		\overline{V} & = \frac{1}{N_{C_k}}\sum_{i \in C_k}V_i \\
		\overline{\omega} & = \frac{1}{N_{C_k}}\sum_{i \in C_k}\omega_i \\
		\overline{\theta} & = \frac{1}{N_{C_k}}\sum_{i \in C_k}\theta_i \\
		\overline{\gamma} & = \frac{1}{N_{C_k}}\sum_{i \in C_k}\gamma_i \\
		\overline{\tau} & = \frac{1}{N_{C_k}}\sum_{i \in C_k}\tau_i
	\end{aligned}
	\right.
\end{equation}
%Aa3#
Where $N_{C_k}$ is number of days in cluster $C_k$.

Based on the clustering results presented in \Tab \ref{result}, the characteristics of dynamic traffic flow ($ \mean{Q}$) and dynamic capacity ($ \mean{V}$) across different categories were analyzed. The analysis indicates that the values of dynamic traffic flow ($ \mean{Q}$) and dynamic capacity ($ \mean{V}$) are relatively close in Cluster $C_1$ and Cluster $C_2$. Specifically, $ \mean{Q}$ equals 61.8139 and $ \mean{V}$ equals 61.4440 for $C_1$, whereas $ \mean{Q}$ is 57.4059 and $ \mean{V}$ is 57.6513 for $C_2$. The capacity-flow relationship within these two categories remains relatively balanced, which provides a fundamental premise for the stable operation of the airport and is consistent with the operational scenario of "demand-supply adaptation and stable operation". Cluster $C_3$ is characterized by a significantly higher dynamic traffic flow than dynamic capacity ($ \mean{Q} = 50.0734$ and $ \mean{V} = 52.7624$), representing a typical scenario of capacity-flow imbalance. Consistently, the average flight on-time rate in this category ($ \mean{\gamma} = 0.5985$) is significantly lower than that in $C_1$ and $C_2$, while the average flight delay time ($ \mean{\tau} = 36.0798$ minutes) is remarkably longer. This indicates that $C_3$ corresponds to a typical flight delay pattern caused by capacity-flow imbalance. Cluster $C_4$ is relatively special: although both dynamic capacity and traffic flow are at low levels ($ \mean{Q} = 40.2491$ and $ \mean{V} = 39.9474$), the relationship between capacity and flow remains relatively balanced. Furthermore, the flight on-time rate in $C_4$ reaches the maximum value ($ \mean{\gamma} = 0.8645$), while the average flight delay time is the shortest ($ \mean{\tau} = 4.9089$ minutes). This phenomenon demonstrates that the air traffic flow management measures have achieved favorable effects.

\begin{figure}[htbp]
	\centering
	\includegraphics[scale=0.4]{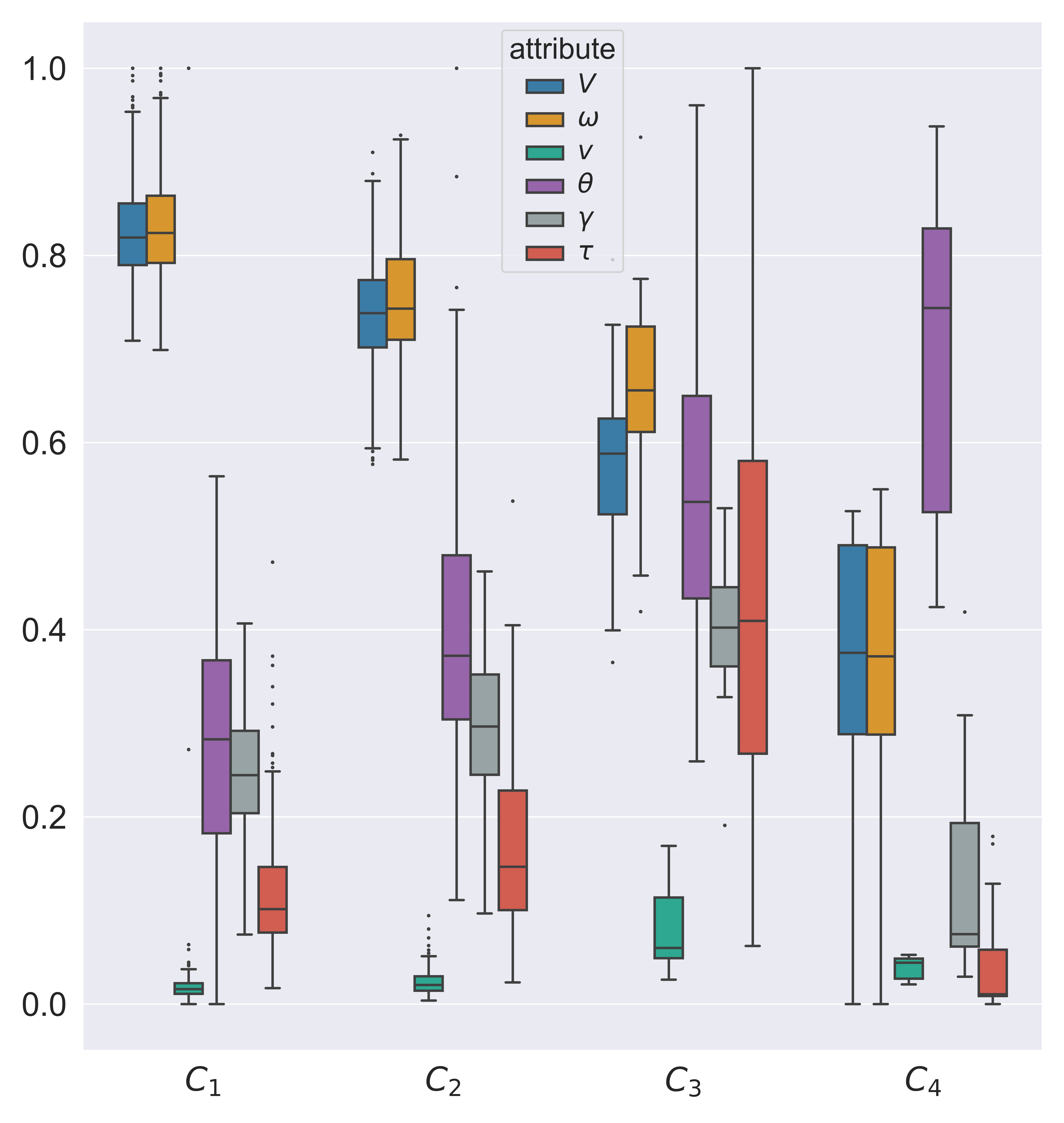}
	\caption{Boxplots of indicators across clusters for Guangzhou Baiyun International Airport}
	\label{CAN_boxplot}
\end{figure}

To further explore the intrinsic mechanisms underlying the relatively high operational efficiency of Clusters \(C_1\), \(C_2\), and \(C_4\), an analysis of their dynamic fluctuation characteristics was conducted. Herein, \(\mean{\omega}\) denotes the dynamic traffic flow change rate and \(\mean{\theta}\) denotes the dynamic capacity change rate. For Cluster \(C_1\), both \(\mean{\omega} = 0.1386\) and \(\mean{\theta} = 0.0851\) are at the lowest levels. This indicates that the fluctuations in dynamic capacity and traffic flow are extremely minimal, and the airport operates in a highly stable state—consistent with the optimal operational characteristics of "low change rate and high stability". Cluster \(C_2\) exhibits slightly lower dynamic capacity and traffic flow compared to Cluster \(C_1\), while its \(\mean{\omega} = 0.1380\) and \(\mean{\theta} = 0.1068\) are marginally higher. This suggests that its operational state is affected by minor disturbances, such as sudden short-term changes in traffic flow. Although its stability is slightly lower than that of Cluster \(C_1\), it remains within a controllable range, which aligns with the regular operational scenario of "dynamic disturbances and reduced stability". For Cluster \(C_4\), both \(\mean{\omega} = 0.1780\) and \(\mean{\theta} = 0.1584\) are higher than those of Clusters \(C_1\) and \(C_2\), indicating frequent adjustments to its dynamic capacity and traffic flow. Combined with its extremely high operational efficiency, it can be inferred that an artificial regulation mechanism exists in this cluster: by proactively adjusting demand and capacity while sacrificing a certain degree of operational scale, efficient and low-delay operation is achieved. The boxplots of Guangzhou Baiyun International Airport, presented in \Fig \ref{CAN_boxplot}, intuitively illustrate the distribution patterns of indicators across different clusters. These boxplots demonstrate that all operational modes are characterized by high flight on-time rates and short delay times, sharing the prominent feature of "dynamic capacity-flow balance". As evidenced by Cluster \(C_4\), such balance can be achieved through manual supervision and control measures.

In conclusion, the operational states of the airport can be categorized into four distinct clusters, as summarized below:
\begin{itemize}
	\item Cluster \(C_1\): High-traffic and high-efficiency operation mode. \\ This cluster is characterized by large dynamic capacity and traffic flow, stable operation, and the highest overall efficiency.
	\item Cluster \(C_2\): Traffic-saturated and moderate congestion mode. \\ The dynamic capacity and traffic flow in this cluster are slightly lower than those in \(C_1\). While it may be affected by short-term traffic fluctuations, its operational efficiency remains within an acceptable range.
	\item Cluster \(C_3\): Traffic-saturated and severe congestion mode. \\
	Although a certain level of dynamic capacity and traffic flow is maintained under this mode, it exhibits extremely poor operational efficiency and significant fluctuations—making it a mode that should be strenuously avoided.
	\item Cluster \(C_4\): Mild congestion mode under traffic control. \\
	This mode achieves high efficiency and low delay by sacrificing operational scale. The fluctuations observed herein are attributed to proactive regulation, thus classifying it as a special optimized operation mode.
\end{itemize}

\subsection{Discussion}
After conducting clustering analysis on Guangzhou Baiyun International Airport from both statistical and temporal dimensions, the results indicate that the airport congestion characteristics proposed in this study can effectively realize the clustering and classification of its operational status. To verify the generalizability of this method, the research scope was further expanded to six additional airports, namely Beijing Capital International Airport(IATA Code: PEK), Beijing Daxing International Airport(IATA Code: PKX), Shanghai Hongqiao International Airport(IATA Code: SHA), Shanghai Pudong International Airport(IATA Code: PVG), Chengdu Shuangliu International Airport(IATA Code: CTU), and Chengdu Tianfu International Airport(IATA Code: TFU). The same type of clustering analysis was performed on these six airports.

\begin{figure}[htbp]
	\centering
	\subfigure[PEK]{\label{ed_PEK}\includegraphics[width=0.3\textwidth]{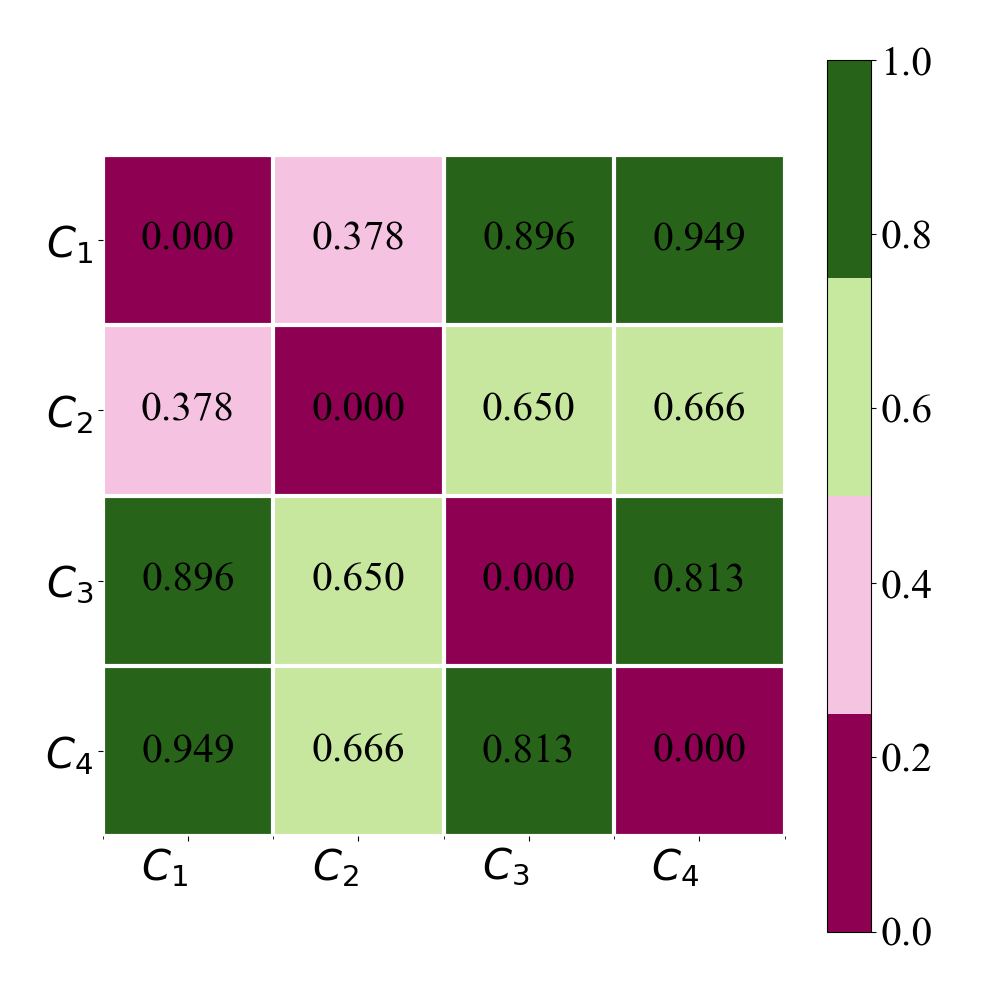}}
	\subfigure[SHA]{\label{ed_SHA}\includegraphics[width=0.3\textwidth]{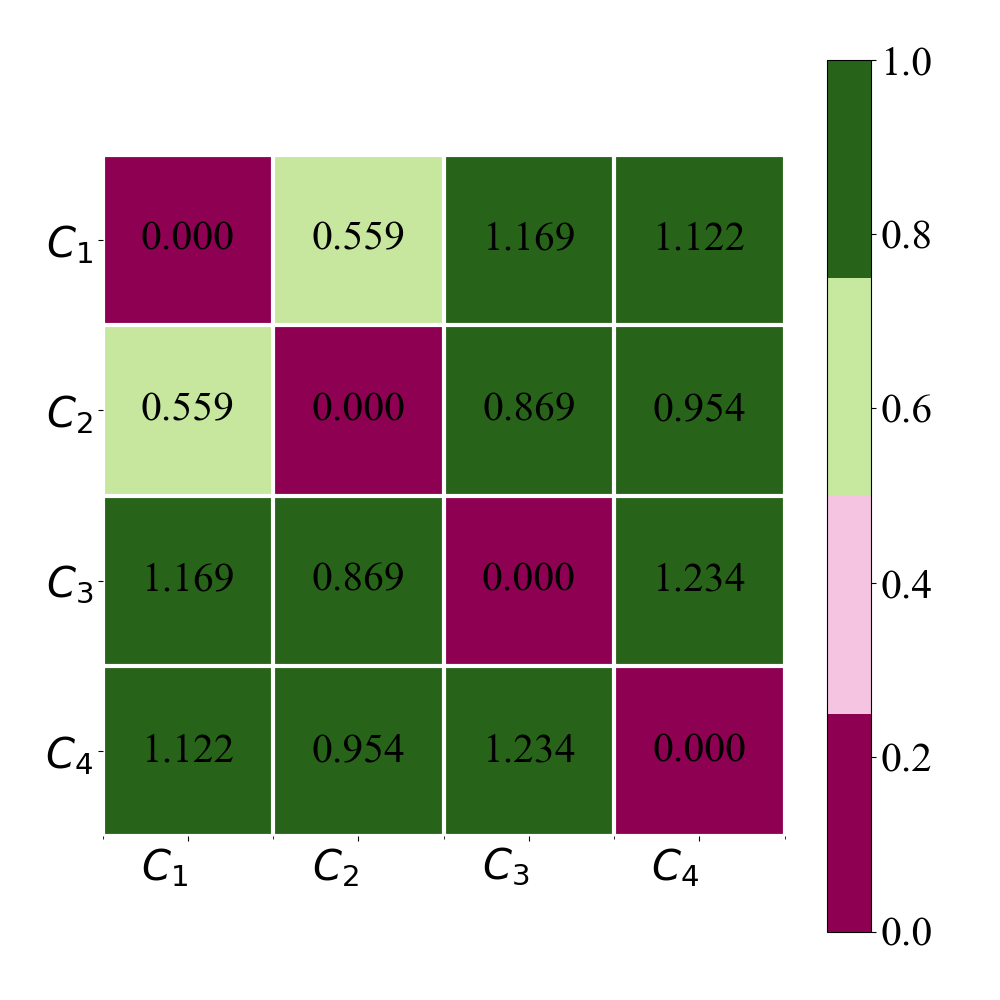}}
	\subfigure[CTU]{\label{ed_CTU}\includegraphics[width=0.3\textwidth]{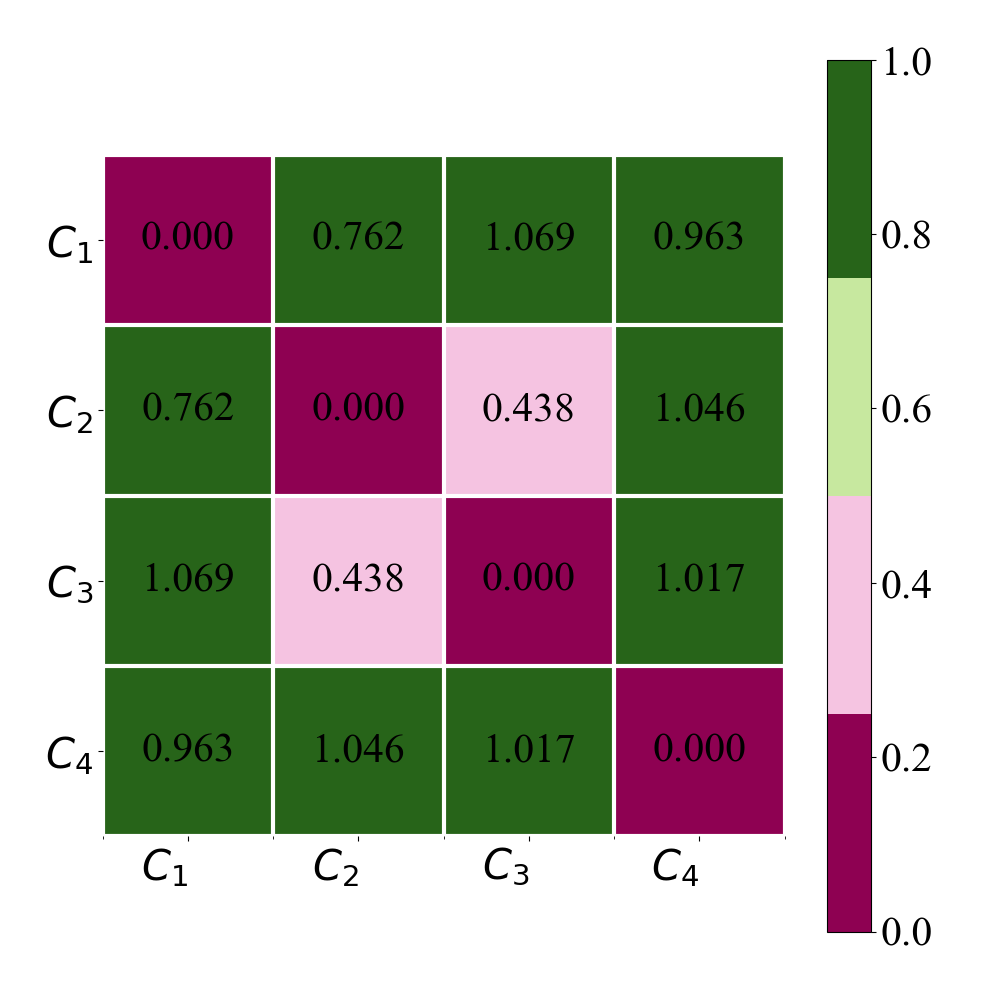}}
	\\
	\subfigure[PKX]{\label{ed_PKX}\includegraphics[width=0.3\textwidth]{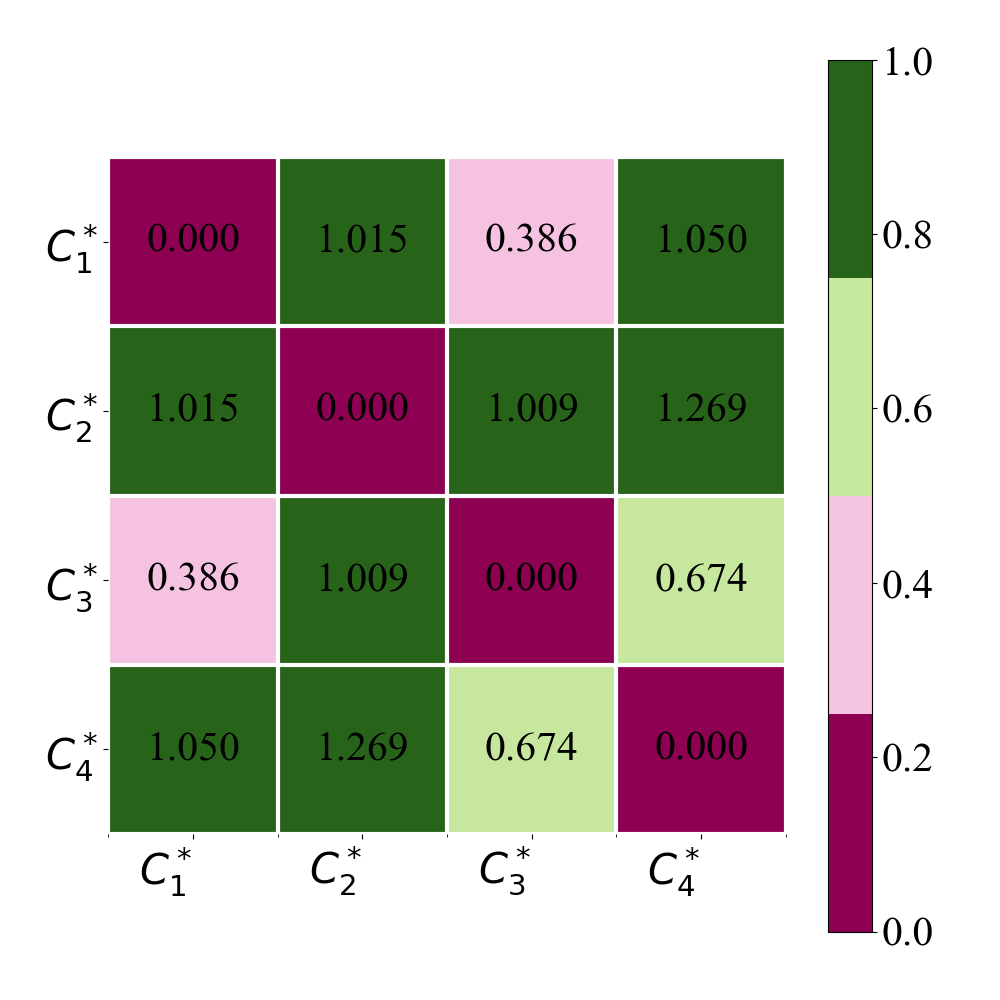}}
	\subfigure[PVG]{\label{ed_PVG}\includegraphics[width=0.3\textwidth]{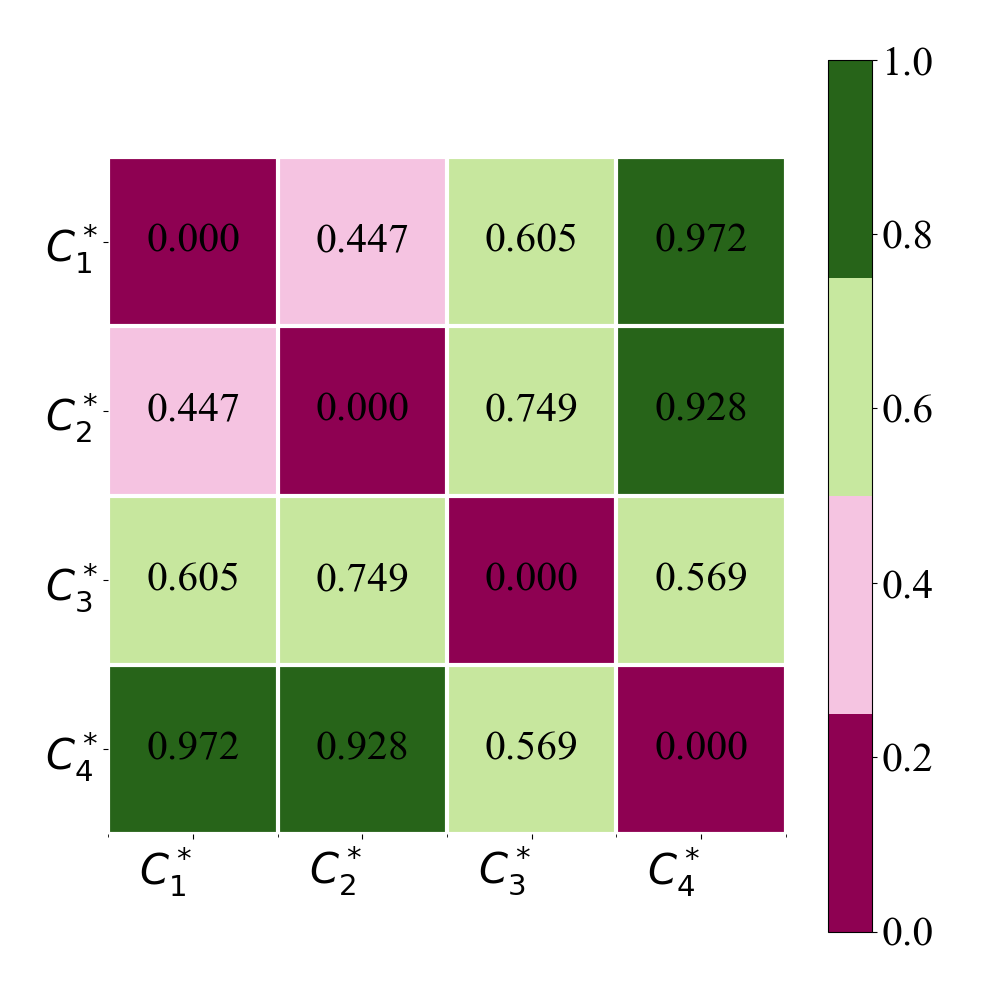}}
	\subfigure[TFU]{\label{ed_TFU}\includegraphics[width=0.3\textwidth]{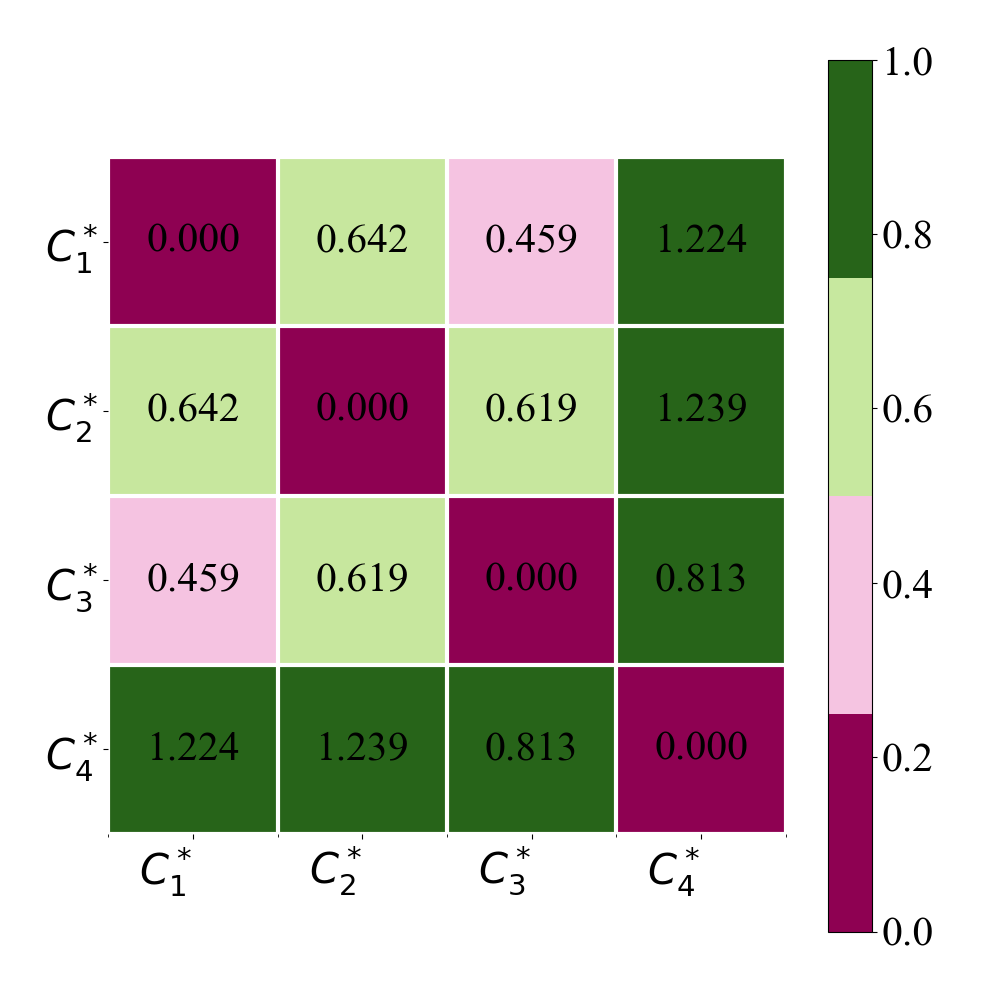}}
	\caption{Euclidean Distance Matrix for Six Airports}
	\label{airports_ed}
\end{figure}

\begin{figure}[htbp]
	\centering
	\subfigure[PEK]{\label{3D_PEK}\includegraphics[width=0.3\textwidth]{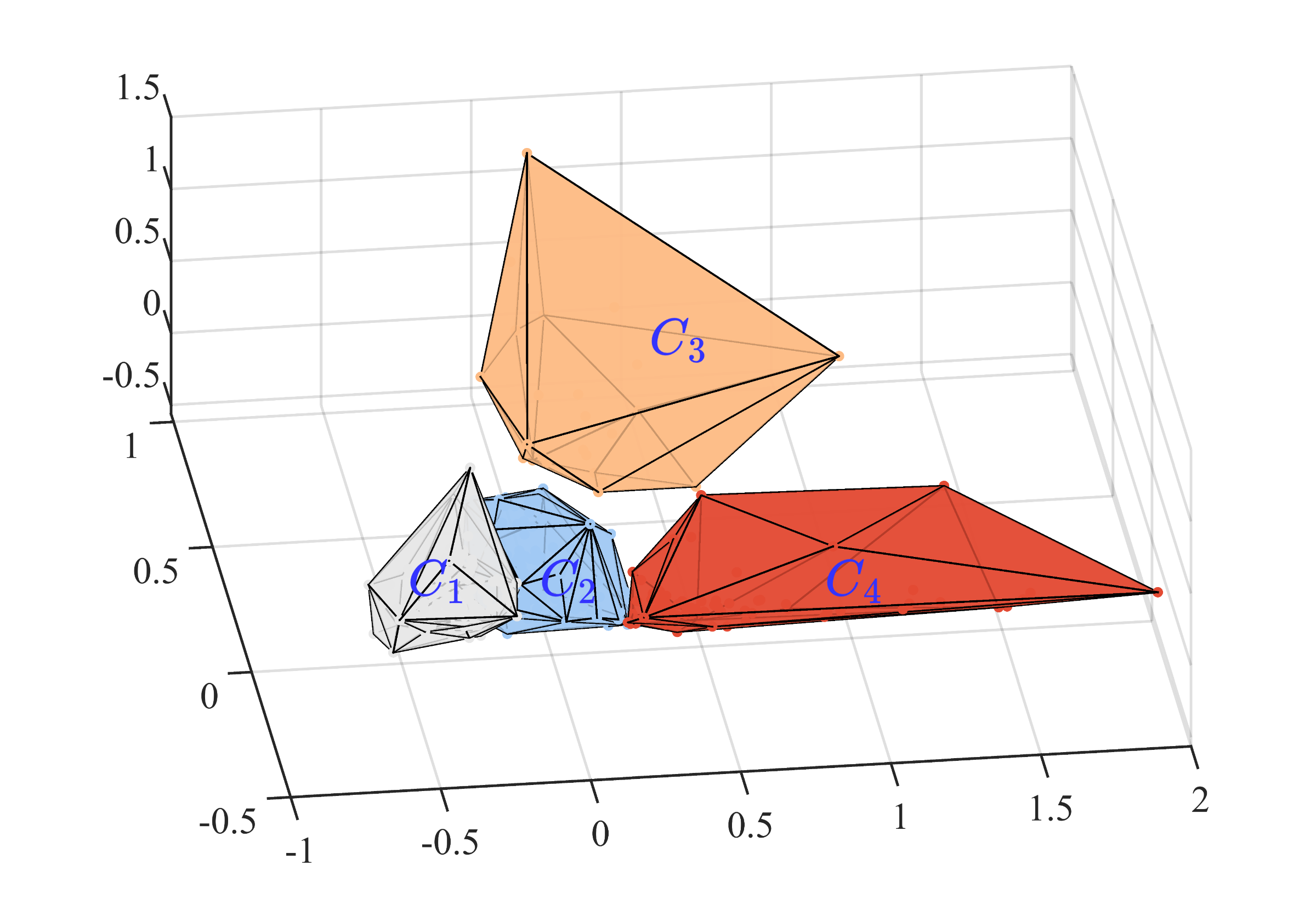}}
	\subfigure[SHA]{\label{3D_SHA}\includegraphics[width=0.3\textwidth]{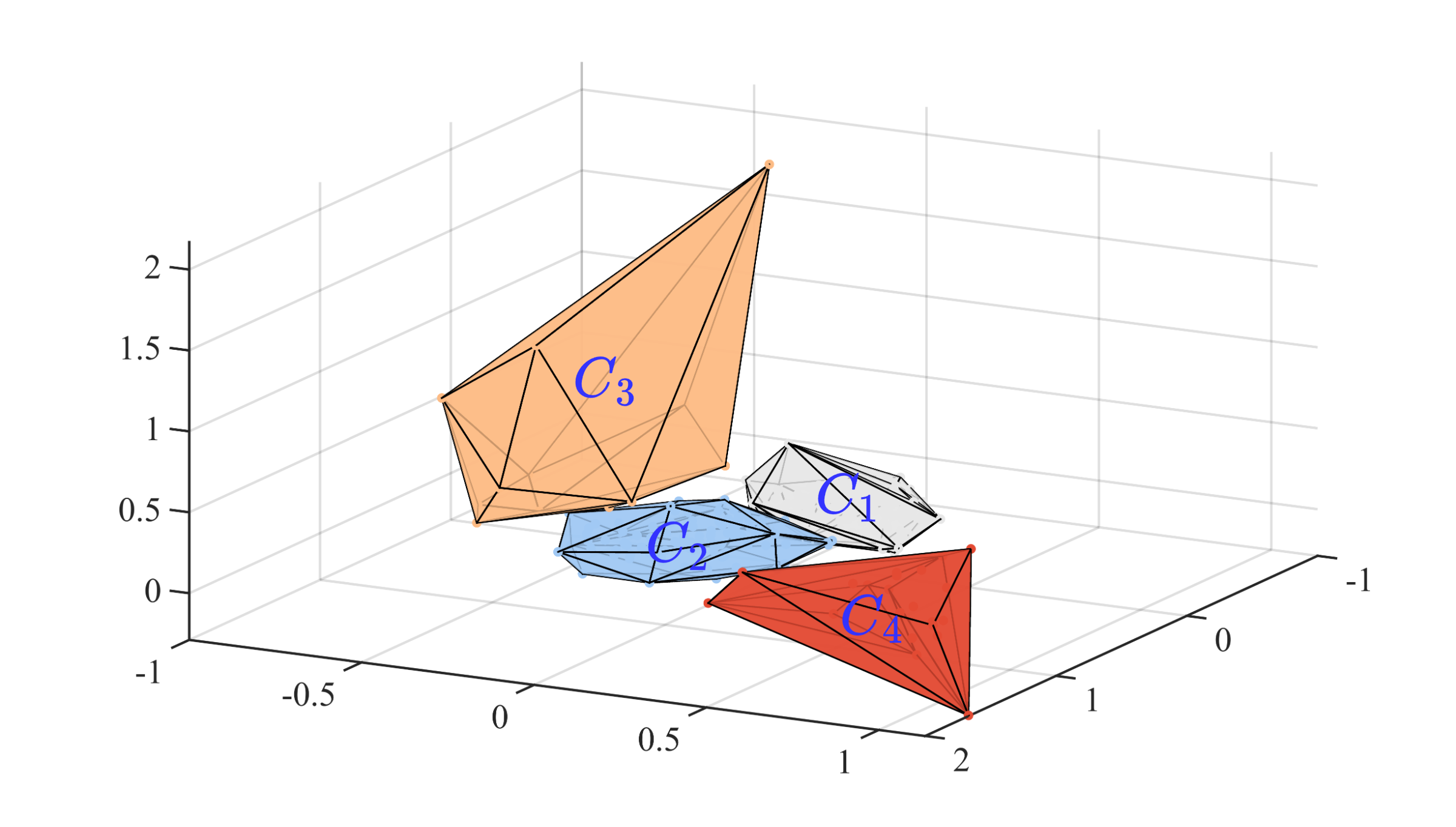}}
	\subfigure[CTU]{\label{3D_CTU}\includegraphics[width=0.3\textwidth]{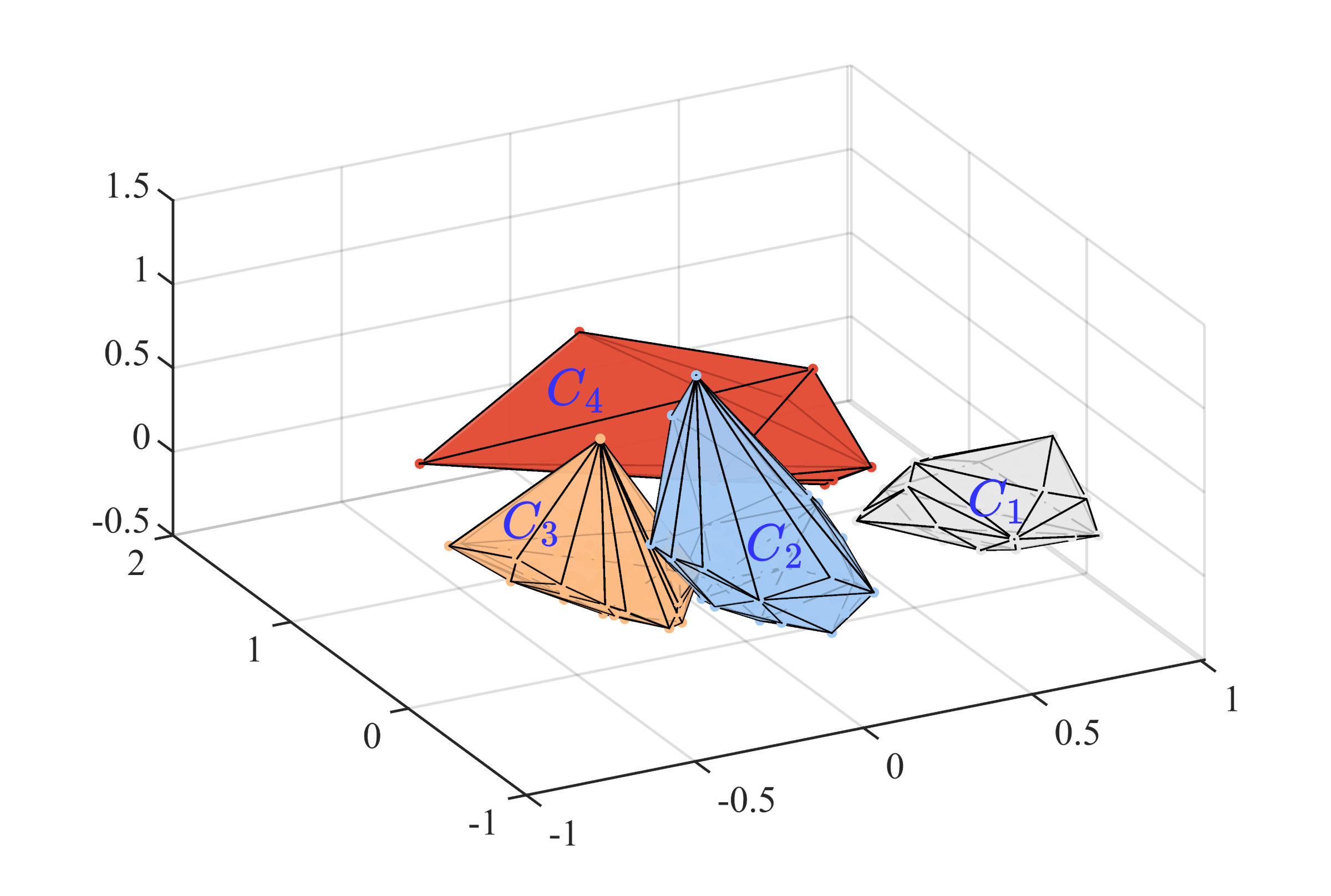}}
	\\
	\subfigure[PKX]{\label{3D_PKX}\includegraphics[width=0.3\textwidth]{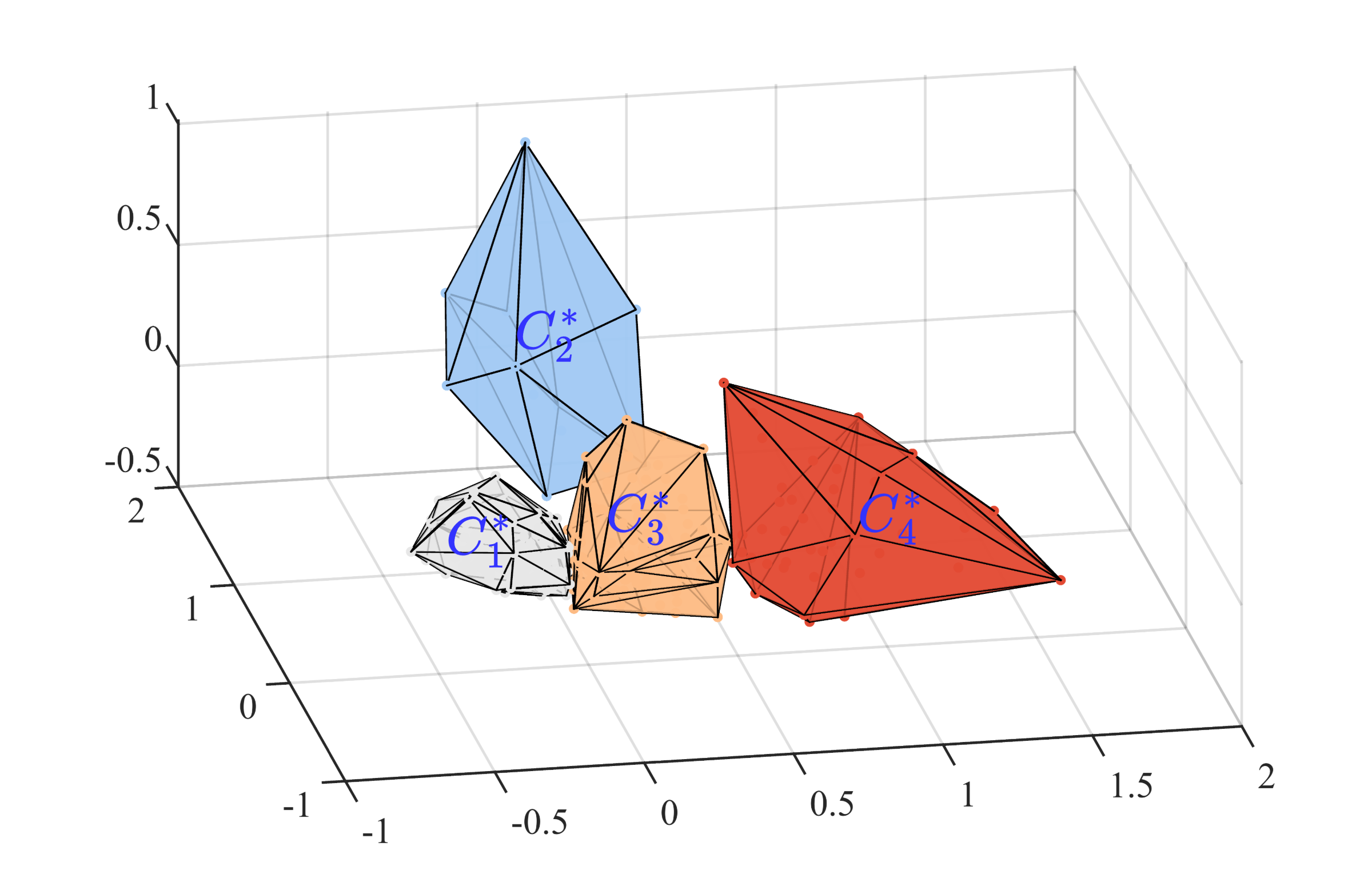}}
	\subfigure[PVG]{\label{3D_PVG}\includegraphics[width=0.3\textwidth]{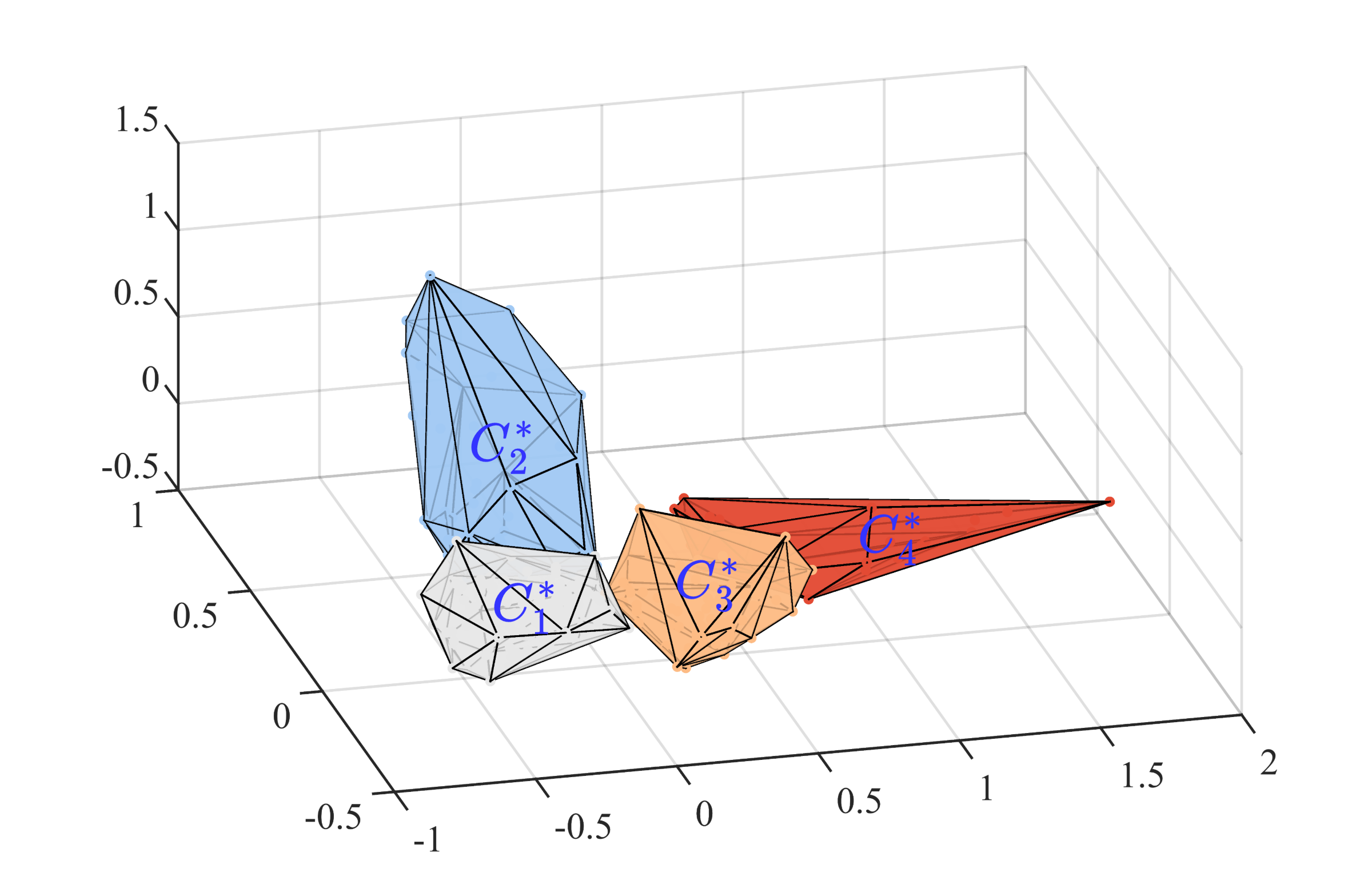}}
	\subfigure[TFU]{\label{3D_TFU}\includegraphics[width=0.3\textwidth]{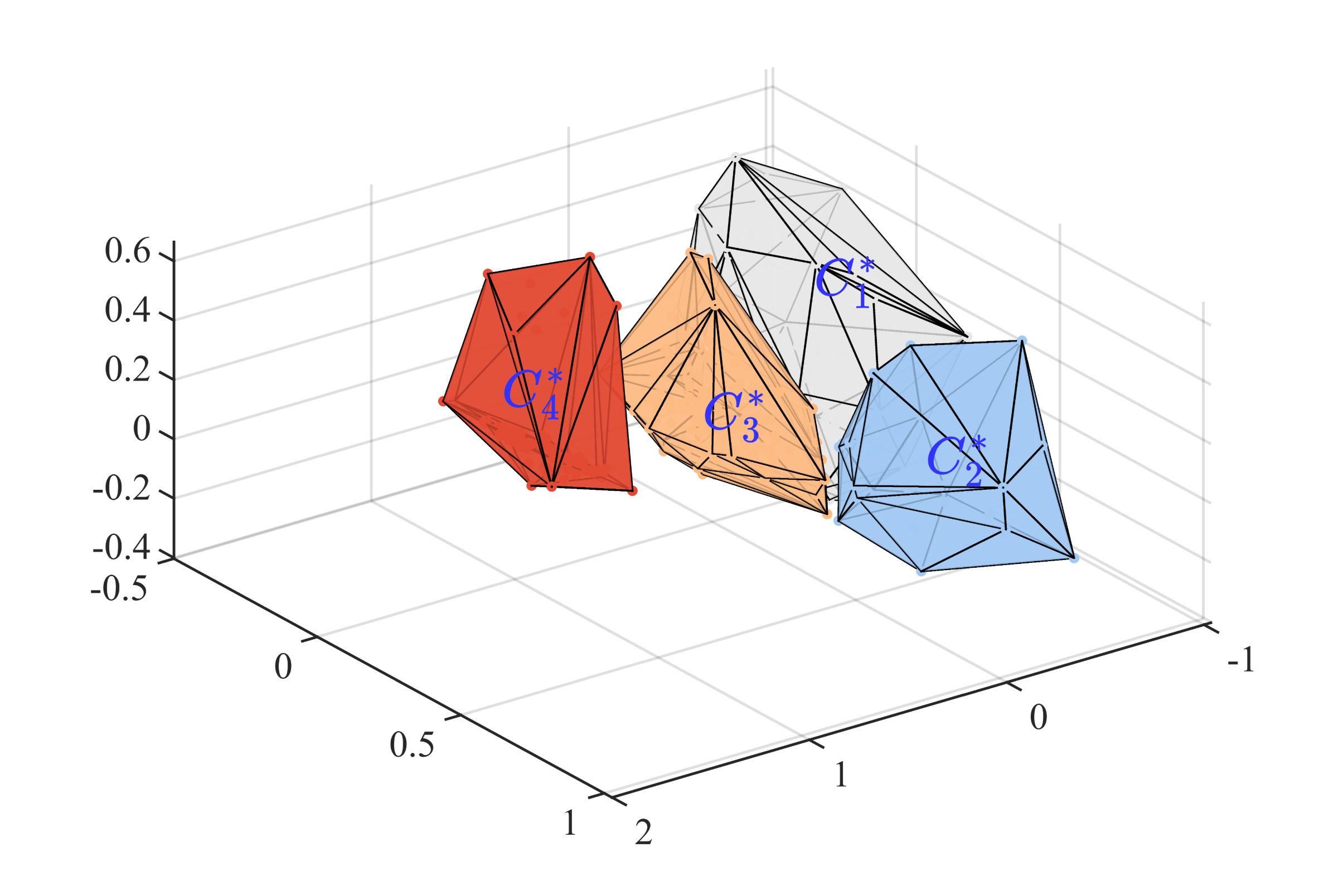}}
	\caption{Visualization of Clustering Results for Six Airports}
	\label{airports_3D_plot}
\end{figure}

The Euclidean distance matrix between the cluster centers of each airport is presented in \Fig \ref{airports_ed}. As observed from the matrix results, the distances between cluster centers of most airports are relatively large, indicating good distinguishability between different categories. However, effective differentiation remains challenging for certain categories—specifically, Clusters \(C_1\) and \(C_2\) at Beijing Capital International Airport and Clusters \(C_2\) and \(C_3\) at Chengdu Shuangliu International Airport—when relying solely on the distance between cluster centers. To address this issue, principal component analysis (PCA) was adopted in this study to reduce the dimensionality of the original data and enable visualization. The optimal display perspective was selected, and the resulting visualization is shown in \Fig \ref{airports_3D_plot}. The visualization results demonstrate that even for categories with relatively close cluster center distances, the clustering algorithm can still achieve clear classification.

It is important to note that the six airports selected in this study share a key common characteristic: they all conform to the "one-city, two-airports" layout model. Specifically, within the same city, there are two airports in operation: an "existing airport" with an extensive operational history and a "newly constructed airport" commissioned in recent years. A typical example is the pairing of Beijing Capital International Airport and Beijing Daxing International Airport. Based on the airport congestion evaluation indicators proposed in Section \ref{sec3.2} of this study, a quantitative evaluation was conducted on the clustering results of the six airports, with the specific evaluation results presented in Tables \ref{PEK_result} - \ref{TFU_result}. During the analysis, a key finding was obtained: the existing and newly constructed airports in the same city exhibit distinctly different operational modes. Accordingly, the operational mode category of the existing airport is defined as \(C\), while that of the newly constructed airport is defined as \(C^*\).

\begin{table}[htbp]
	\caption{Clustering Evaluation Results for Beijing Capital International Airport}
	\centering
	\begin{tabular}{llllllll}
		\toprule
		$C_k$ & $\overline{Q}$ & $\overline{V}$ & $\overline{\omega}$ & $\overline{\theta}$ & $\overline{\gamma}$ & $\overline{\tau}$ & Days \\
		\midrule
		$C_1$ & 61.8139    & 61.4440    & 0.1386        & 0.0851        & 0.7502        & 7.9863       & 181 \\
		$C_2$ & 57.4059    & 57.6513    & 0.1380        & 0.1068        & 0.7050        & 12.1177      & 136 \\
		$C_3$ & 50.0734    & 52.7624    & 0.2778        & 0.1360        & 0.5985        & 36.0798      & 33  \\
		$C_4$ & 40.2491    & 39.9474    & 0.1780        & 0.1584        & 0.8645        & 4.9089       & 15  \\
		\bottomrule
	\end{tabular}
\label{PEK_result}
\end{table}

\begin{table}[htbp]
	\caption{Clustering Evaluation Results for Shanghai Hongqiao International Airport}
	\centering
	\begin{tabular}{llllllll}
		\toprule
		$C_k$ & $\overline{Q}$ & $\overline{V}$ & $\overline{\omega}$ & $\overline{\theta}$ & $\overline{\gamma}$ & $\overline{\tau}$ & Days \\
		\midrule
		$C_1$ & 38.2975    & 38.2996    & 0.2390        & 0.1682        & 0.8207        & 5.8184       & 225 \\
		$C_2$ & 38.0705    & 38.2795    & 0.2281        & 0.1735        & 0.7482        & 13.7257      & 106 \\
		$C_3$ & 36.0121    & 37.3765    & 0.4663        & 0.1908        & 0.6312        & 34.5212      & 13  \\
		$C_4$ & 29.4737    & 29.4862    & 0.2556        & 0.2304        & 0.8906        & 2.7854       & 21  \\
		\bottomrule
	\end{tabular}
\label{SHA_result}
\end{table}

\begin{table}[htbp]
	\caption{Clustering Evaluation Results for Chengdu Shuangliu International Airport}
	\centering
	\begin{tabular}{llllllll}
		\toprule
		$C_k$ & $\overline{Q}$ & $\overline{V}$ & $\overline{\omega}$ & $\overline{\theta}$ & $\overline{\gamma}$ & $\overline{\tau}$ & Days \\
		\midrule
		$C_1$ & 29.6924    & 29.5686    & 0.3062        & 0.2433        & 0.8835        & 3.1249       & 71  \\
		$C_2$ & 28.3312    & 28.4553    & 0.1864        & 0.1231        & 0.8081        & 8.7634       & 140 \\
		$C_3$ & 26.5756    & 26.5068    & 0.1781        & 0.0944        & 0.6948        & 10.0065      & 143 \\
		$C_4$ & 25.3541    & 25.2344    & 0.4229        & 0.2955        & 0.9315        & 2.4151       & 11  \\
		\bottomrule
	\end{tabular}
\label{CTU_result}
\end{table}

\begin{table}[htbp]
	\caption{Clustering Evaluation Results for Beijing Daxing International Airport}
	\centering
	\begin{tabular}{llllllll}
		\toprule
		$C_k$  & $\overline{Q}$ & $\overline{V}$ & $\overline{\omega}$ & $\overline{\theta}$ & $\overline{\gamma}$ & $\overline{\tau}$ & Days \\
		\midrule
		$C_1^*$ & 44.2804    & 44.4801    & 0.4469        & 0.3670        & 0.8404        & 6.9450       & 189 \\
		$C_2^*$ & 39.0088    & 41.3728    & 0.6059        & 0.3838        & 0.6360        & 44.2350      & 12  \\
		$C_3^*$ & 39.2640    & 39.3477    & 0.6687        & 0.5732        & 0.8670        & 4.9125       & 122 \\
		$C_4^*$ & 31.0113    & 30.9336    & 0.9858        & 1.2530        & 0.8974        & 4.0173       & 42  \\
		\bottomrule
	\end{tabular}
\label{PKX_result}
\end{table}

\begin{table}[htbp]
	\caption{Clustering Evaluation Results for Shanghai Pudong International Airport}
	\centering
	\begin{tabular}{llllllll}
		\toprule
		$C_k$  & $\overline{Q}$ & $\overline{V}$ & $\overline{\omega}$ & $\overline{\theta}$ & $\overline{\gamma}$ & $\overline{\tau}$ & Days \\
		\midrule
		$C_1^*$ & 60.7877    & 60.7441    & 0.1697        & 0.1158        & 0.7164        & 9.6628       & 146 \\
		$C_2^*$ & 58.7903    & 60.1068    & 0.1726        & 0.1136        & 0.6191        & 19.8313      & 67  \\
		$C_3^*$ & 47.9513    & 47.8408    & 0.2105        & 0.1224        & 0.7338        & 8.5284       & 80  \\
		$C_4^*$ & 40.3209    & 40.3472    & 0.2125        & 0.1759        & 0.7875        & 6.1311       & 72  \\
		\bottomrule
	\end{tabular}
\label{PVG_result}
\end{table}

\begin{table}[htbp]
	\caption{Clustering Evaluation Results for Chengdu Tianfu International Airport}
	\centering
	\begin{tabular}{llllllll}
		\toprule
		$C_k$  & $\overline{Q}$ & $\overline{V}$ & $\overline{\omega}$ & $\overline{\theta}$ & $\overline{\gamma}$ & $\overline{\tau}$ & Days \\
		\midrule
		$C_1^*$ & 51.3428    & 51.4426    & 1.0100        & 0.3669        & 0.7975        & 7.9621       & 115 \\
		$C_2^*$ & 50.1885    & 51.1969    & 0.8063        & 0.3373        & 0.6757        & 22.8506      & 31  \\
		$C_3^*$ & 47.0425    & 47.0846    & 0.6202        & 0.3493        & 0.8228        & 6.2311       & 135 \\
		$C_4^*$ & 36.5501    & 36.5219    & 0.3170        & 0.1903        & 0.8379        & 3.3835       & 84  \\
		\bottomrule
	\end{tabular}
\label{TFU_result}
\end{table}

\begin{figure}[htbp]
	\centering
	\subfigure[PEK]{\label{box_PEK}\includegraphics[width=0.3\textwidth]{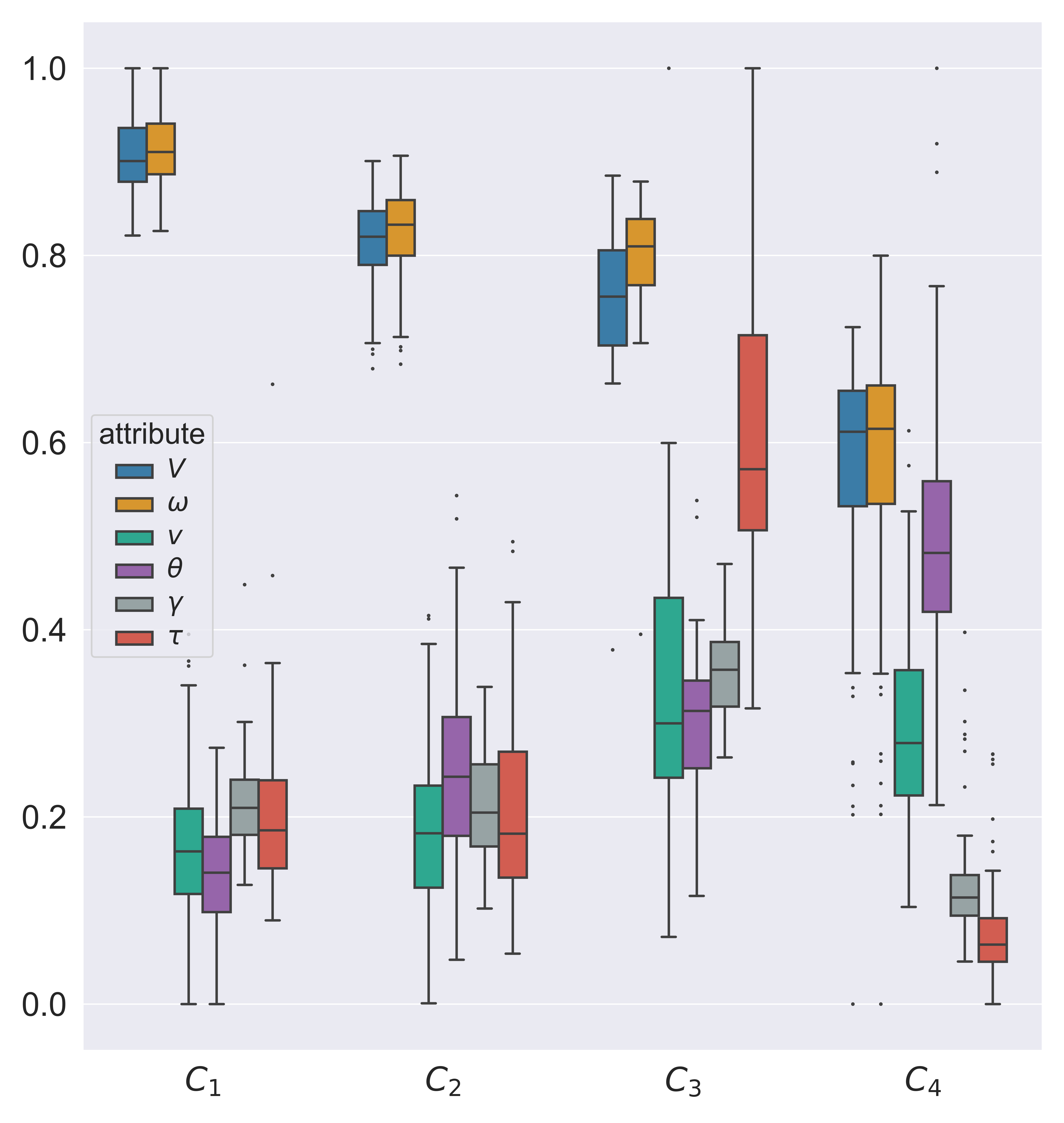}}
	\subfigure[SHA]{\label{box_SHA}\includegraphics[width=0.3\textwidth]{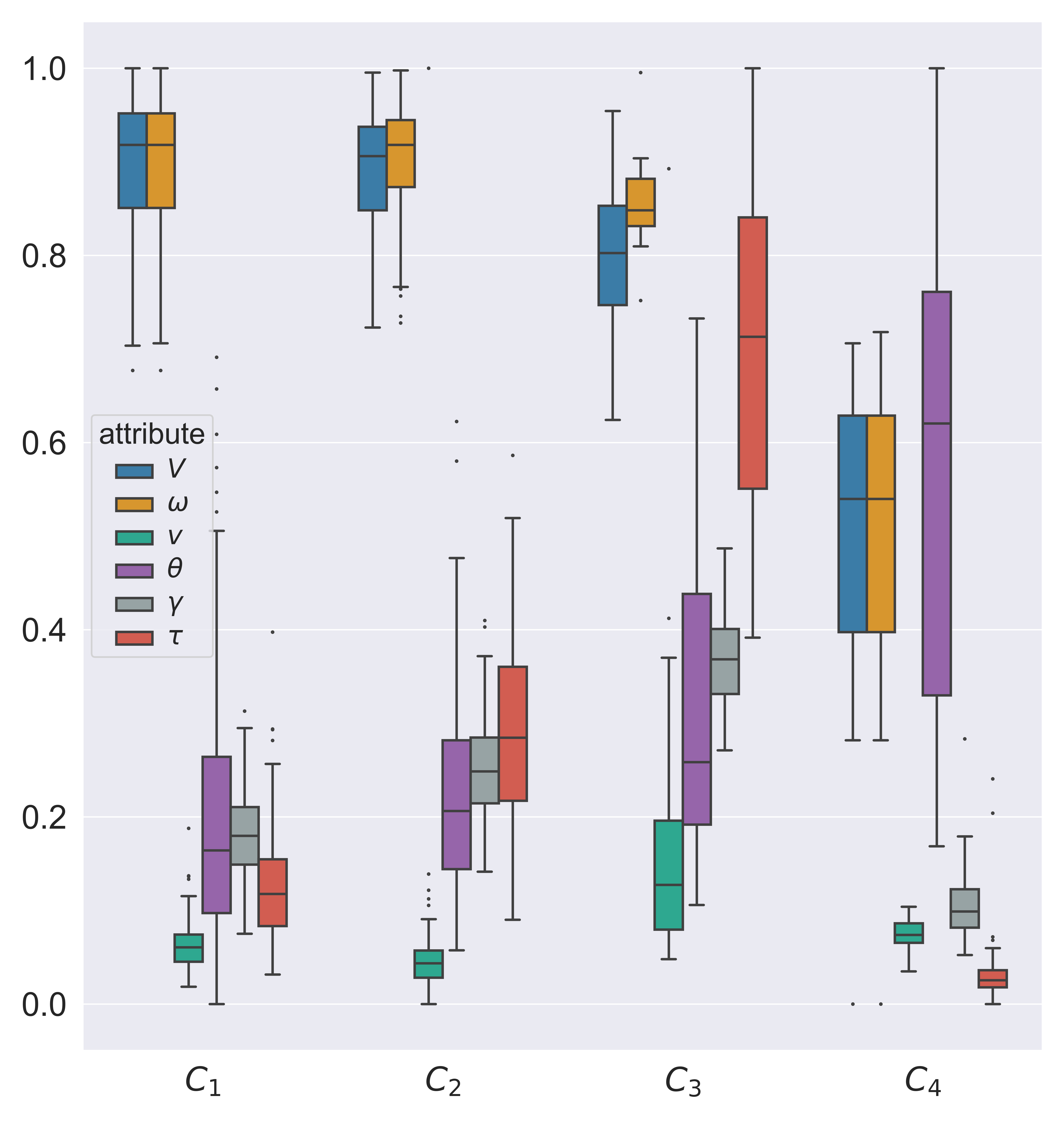}}
	\subfigure[CTU]{\label{box_CTU}\includegraphics[width=0.3\textwidth]{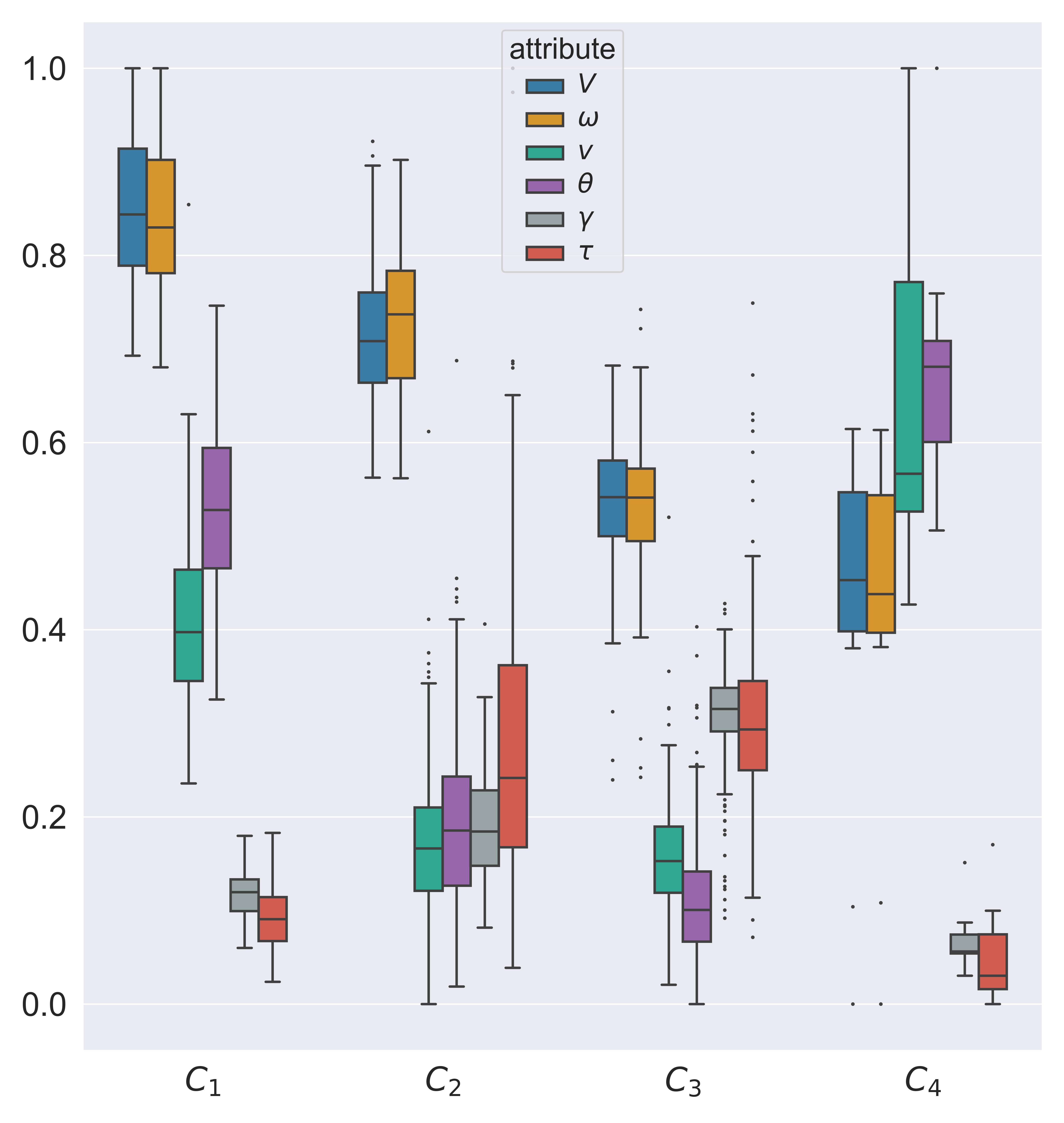}}
	\\
	\subfigure[PKX]{\label{box_PKX}\includegraphics[width=0.3\textwidth]{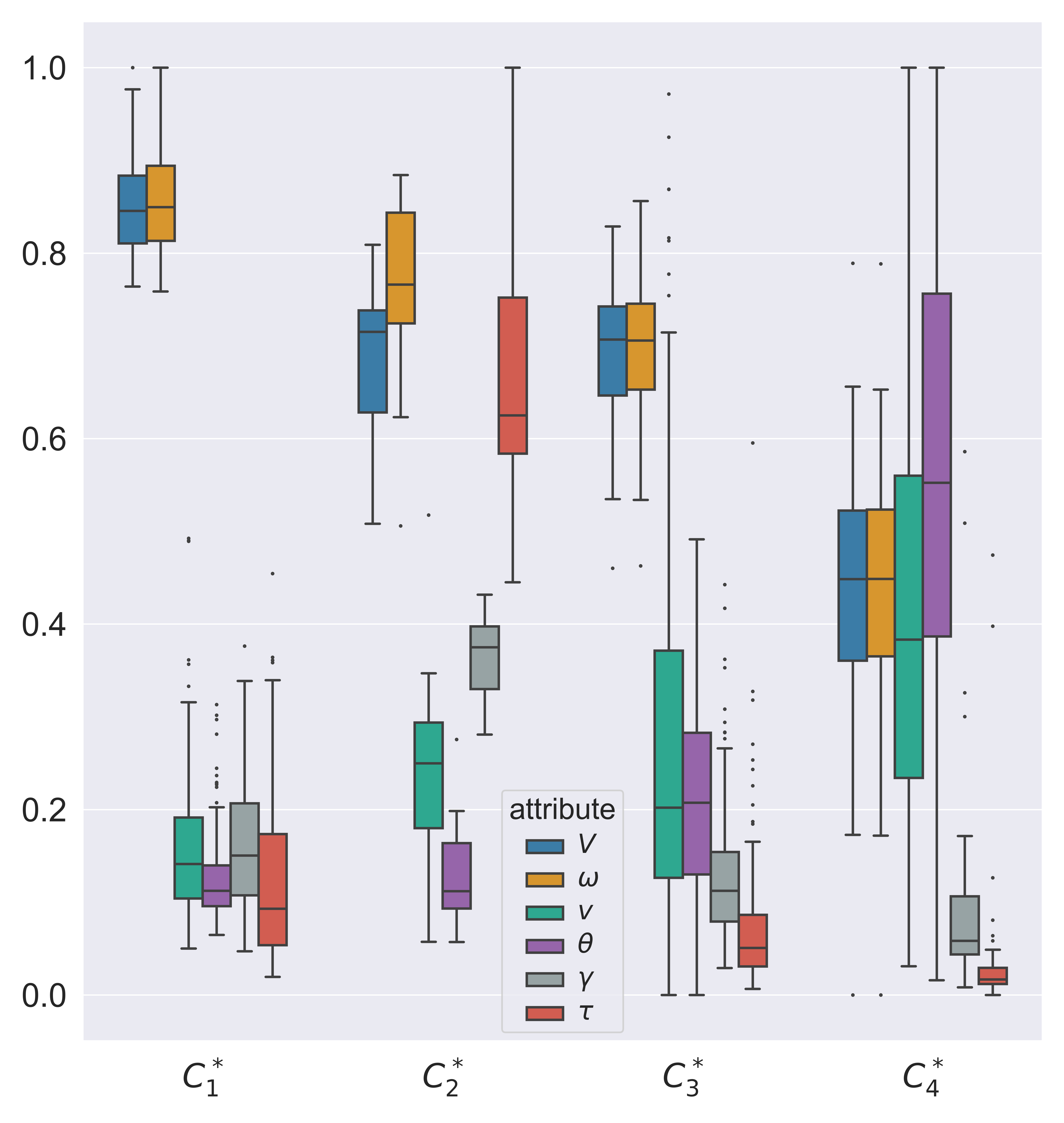}}
	\subfigure[PVG]{\label{box_PVG}\includegraphics[width=0.3\textwidth]{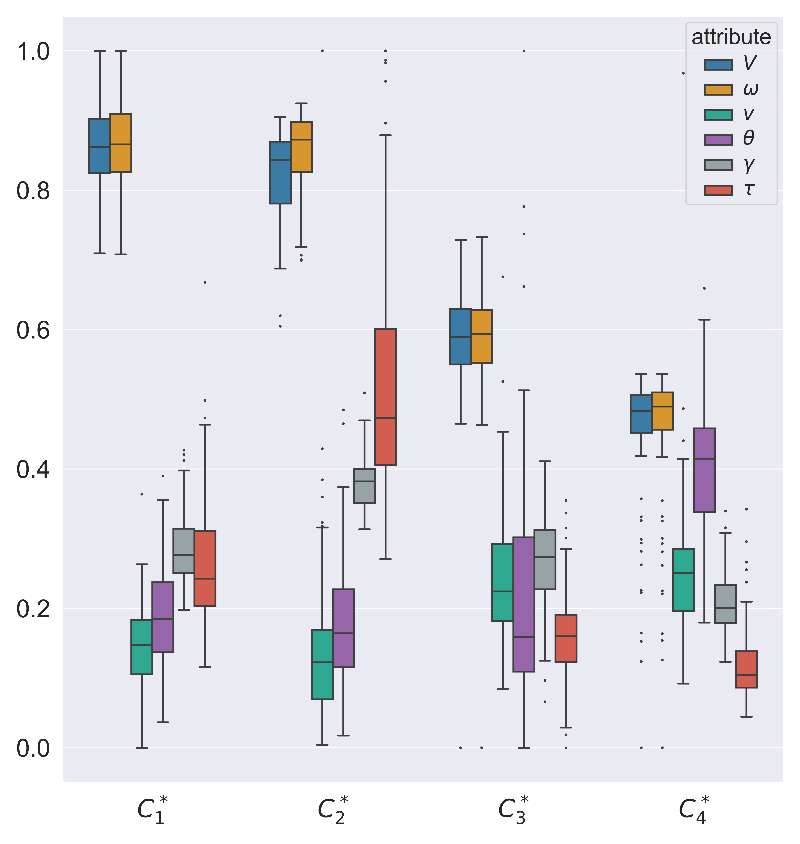}}
	\subfigure[TFU]{\label{box_TFU}\includegraphics[width=0.3\textwidth]{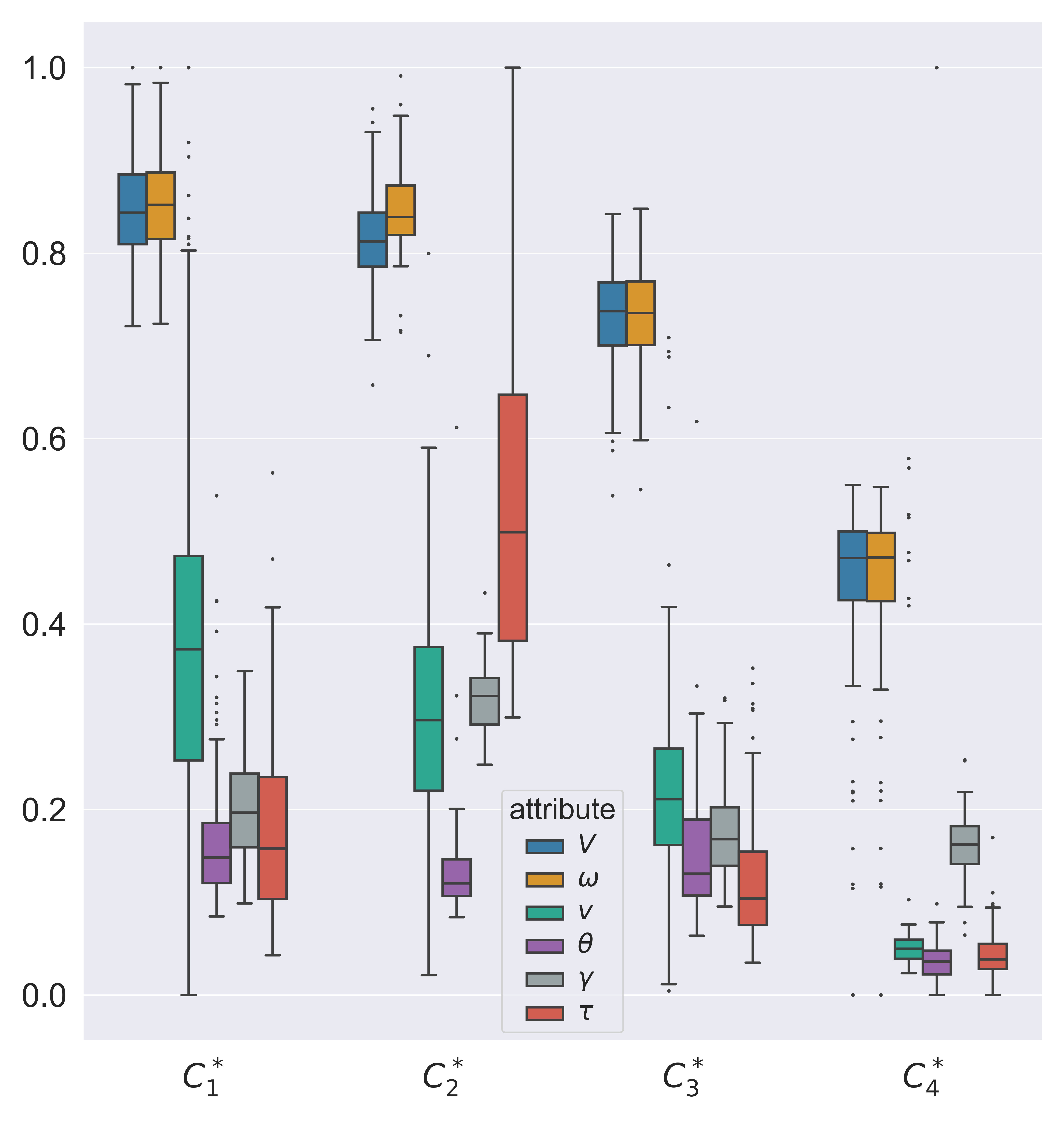}}
	\caption{Boxplots of indicators across clusters for Six Airports}
	\label{airports_box}
\end{figure}

For the existing airports in cities (e.g., Beijing Capital International Airport, Chengdu Shuangliu International Airport, Shanghai Hongqiao International Airport), their operational patterns are consistent with that of Guangzhou Baiyun International Airport as described earlier, with the relevant analysis results presented in \Fig \ref{airports_box}(a) - (c). From the clustering results, these airports exhibit relatively prominent congestion characteristics and operational regularities in Clusters \(C_1\) - \(C_3\), which are specifically manifested as follows: while the flight volume gradually decreases, the flight irregularity rate shows a continuous upward trend, and the average flight delay duration extends synchronously. Essentially, this phenomenon reflects the increasingly prominent contradiction between airport operational resources (e.g., runways, parking stands) and the demand for flight slots.

A further observation of the clustering data for the \(C_1\) - \(C_3\) modes reveals that, in terms of the change rate between the actual hourly flight takeoff/landing movements and the scheduled ones, the fluctuation range gradually expands. This characteristic indicates that during the operation of existing airports, due to the rigid constraints of hardware resources such as runway capacity and the number of parking stands, when external interferences (e.g., adverse weather, airspace control) occur and require adjustments to flight schedules, it is highly likely to cause significant deviations between the actual and scheduled takeoff/landing movements. As a result, the airport can hardly operate stably in accordance with the established flight timetable, which further disrupts the overall operational order. The changing trends of the aforementioned indicators (including flight volume, flight irregularity rate, and average delay duration) are highly consistent with the actual operational conditions of the above-mentioned existing airports. This further confirms that the contradiction between the demand for flight slot resources and airport support resources has become increasingly acute, serving as the core factor restricting the improvement of operational efficiency of existing airports. Notably, when special circumstances such as severe weather or airspace activity control occur, the operational status of the aforementioned existing airports is affected more significantly. In such cases, air traffic flow management measures are usually implemented to substantially reduce the flight volume. From the perspective of operational results, although the total number of flights decreases significantly, the flight on-time rate increases obviously and the average delay duration shortens remarkably. This suggests that air traffic flow management measures can effectively alleviate the operational pressure of airports and improve operational quality in the short term. Although the reduction in flight volume reduces the transportation capacity of airports to a certain extent, an analysis from the dimension of overall operational efficiency shows that this measure helps to improve the flight on-time rate, ensure the smoothness of passengers’ travel, and thus achieves a short-term balance between "prioritizing operational efficiency" and "guaranteeing passenger experience".

The clustering results of Beijing Daxing International Airport, Shanghai Pudong International Airport, and Chengdu Tianfu International Airport exhibit significant differences from those of the aforementioned airports, with their clustering evaluation results presented in Tables \ref{PEK_result} - \ref{TFU_result}. This study takes Beijing Daxing International Airport as a typical case for in-depth analysis. For Cluster \(C_1^*\), it features the highest dynamic capacity-flow volume and the lowest dynamic capacity-flow change rate. This characteristic indicates that the airport operates in a stable and efficient state on the days corresponding to this category. Meanwhile, under this category, the flight regularity rate remains at a high level and the average flight delay duration is maintained at a low level. Therefore, Cluster \(C_1^*\) can be defined as the high-capacity-flow and high-efficiency operation mode of Beijing Daxing International Airport. Compared with Cluster \(C_1^*\), Cluster \(C_2^*\) still maintains a relatively high range of dynamic capacity-flow volume, but there is a significant gap between its dynamic capacity-flow volume and that of \(C_1^*\). More critically, under this category, the flight regularity rate is significantly lower and the average flight delay duration is extremely high—representing a high-risk state that requires key avoidance in airport operations. Thus, Cluster \(C_2^*\) can be defined as the traffic-saturated severe congestion mode caused by capacity-flow imbalance. The core characteristics of Categories \(C_3^*\) and \(C_4^*\) are high volatility in dynamic capacity-flow volume, with the volatility of \(C_4^*\) being particularly prominent. Notably, both categories exhibit extremely high flight regularity rates and extremely low average flight delay durations—even outperforming the flight efficiency indicators of \(C_1^*\) under the "optimal-efficiency regular operation" state. The key difference between the two lies in the magnitude of dynamic capacity-flow volume: Category \(C_3^*\) is characterized by a relatively high dynamic capacity-flow volume, corresponding to the "moderate congestion under traffic flow management" operational mode; while Category \(C_4^*\) has a low dynamic capacity-flow volume, corresponding to the "mild congestion under traffic flow management" operational mode.It is worth noting that Clusters \(C_3^*\) and \(C_2^*\) have similar magnitudes of dynamic capacity-flow volume but drastically different flight operational states. This implies that even when the airport is in different complex scenarios, the congestion modes under traffic control can not only ensure the quantity of flight operations but also guarantee the quality of flight execution. Even in extreme operational conditions (e.g., Category \(C_4^*\)), the quality of flight execution can be ensured by sacrificing the quantity of flight operations. The clustering evaluation results for Beijing Daxing International Airport, Shanghai Pudong International Airport, and Chengdu Tianfu International Airport are intuitively visualized using boxplots, as presented in \Fig \ref{airports_box}(d) - (f). From the figures, it can be clearly observed that the overall operational state distribution of these three airports is highly consistent with the conclusions of the single-case analysis of Beijing Daxing International Airport above.

A further comparison of existing and newly constructed airports in the same city reveals that they exhibit distinctly different operational characteristics: newly constructed airports have sufficient allocation of support resources, leading to an inherent difference in their operational patterns from existing airports. This difference is specifically manifested as follows: when facing similar flight volume adjustment demands, newly constructed airports can still maintain a high level of operational quality; once the flight volume is reduced, the supply-demand contradiction between slot resources and support capacity is quickly alleviated, which in turn promotes a significant improvement in flight regularity and a substantial reduction in average delay duration. The core reason for this advantage lies in the fact that newly constructed airports retain a relatively large redundancy in key infrastructure and support systems, such as runway capacity, the number of parking stands, and air traffic control command systems. This redundancy enables newly constructed airports to more flexibly respond to short-term fluctuations and long-term adjustments in flight volume.

In general, under the "one-city, two-airports" layout, the operational modes of existing airports can be classified into four categories, namely:
\begin{itemize}
	\item High-traffic and high-efficiency operation mode;
	\item Traffic-saturated moderate congestion mode;
	\item Traffic-saturated severe congestion mode;
	\item Mild congestion mode under traffic control.
\end{itemize}
In contrast, the operational modes of newly constructed airports under the same layout can also be categorized into four types, specifically including:
\begin{itemize}
	\item High-traffic and high-efficiency operation mode;
	\item Traffic-saturated moderate congestion mode;
	\item Moderate congestion mode under traffic control;
	\item Mild congestion mode under traffic control.
\end{itemize}
Notably, for most cities in China adopting the "single-airport" layout, the operational mode of their airports is highly consistent with that of the existing airport under the "one-city, two-airports" layout.

\section{Conclusions}

This study focuses on the feature representation and clustering of airport congestion, proposing a time series analysis method for airport congestion that combines the Hurst Exponent with higher-order statistics. The aim is to identify airport congestion patterns through data mining techniques and provide a scientific basis for air traffic management. Based on the 2023 flight data of China's civil aviation industry, an empirical analysis was conducted on a total of 7 airports, including Guangzhou Baiyun International Airport and airports in Beijing, Shanghai, Chengdu, etc. The key conclusions are as follows:

First, the feature extraction and clustering framework proposed in this study exhibits effectiveness and universality. By using 1st to 4th order higher-order cumulants to eliminate the interference of Gaussian random noise, and combining the Hurst Exponent to characterize the long-term correlation and stability of time series, the constructed high-dimensional feature vector can accurately capture the dynamic laws of airport congestion. The K-means clustering algorithm was applied to analyze the data of Guangzhou Baiyun International Airport. The results show that the intra-cluster compactness and inter-cluster separation of the 4 congestion patterns all meet statistical requirements; after dimensionality reduction and visualization via Principal Component Analysis (PCA), each category can be effectively distinguished. When this framework was extended to the other 6 airports, the Euclidean distances between the cluster centers of most airports were relatively large. Even for some categories with relatively close cluster center distances, clear classification could still be achieved through PCA visualization, fully demonstrating the universality of this method.

Second, there are significant differences in the operation modes between existing and newly constructed airports under the "one-city, two-airports" layout. The four operation states of existing airports are: high-traffic and high-efficiency operation mode, traffic-saturated moderate congestion mode, traffic-saturated severe congestion mode, and mild congestion mode under traffic control. Restricted by the rigid constraints of hardware resources such as runways and parking stands, the $C_1$ - $C_3$ operation modes of these existing airports exhibit the characteristic of "decreasing traffic volume - deteriorating efficiency", while the $C_4$ mode shows the characteristic of "high operational fluctuation - improved efficiency". The operation states of newly constructed airports are categorized as: high-traffic and high-efficiency operation mode, traffic-saturated severe congestion mode, moderate congestion mode under traffic control, and mild congestion mode under traffic control. Among them, the $C_1^*$ and $C_2^*$ operation modes also present the "decreasing traffic volume - deteriorating efficiency" characteristic; however, the $C_3^*$ and $C_4^*$ modes rely more on manual regulation, ensuring flight operation efficiency by appropriately sacrificing the number of flights handled by the airport. In addition, the operation mode of most single-airport cities in China is highly consistent with that of existing airports under the "one-city, two-airports" layout, all facing the problem of "increasing contradiction between resource supply and demand".

This study provides differentiated practical strategies for airport congestion governance. For existing airports under the "one-city, two-airports" layout and most single-airport cities in China, the $C_4$ "mild congestion under traffic control" mode can be referenced in the short term: by adjusting flight schedules to avoid peak hours and limiting the number of takeoffs and landings per hour, the on-time rate can be improved at the cost of a small reduction in traffic volume. In the long term, it is necessary to break through hardware constraints by upgrading infrastructure and optimizing airspace structure. For newly constructed airports, full use should be made of infrastructure redundancy, and the "proactive regulation" mechanism centered on the $C_3^*$ mode should be promoted to balance transportation scale and operation efficiency. These clustering results not only reveal the potential problems and challenges in airport operation but also provide valuable guidance for decision-makers, contributing to the sustainable development and scientific management of the air transportation system. The research conclusions can provide decision support for flight schedule optimization, air traffic management, air transportation policy formulation, and early warning system prediction in the civil aviation field. They hold significant theoretical significance and practical application value for improving flight delay issues, and also offer a new perspective for airport management departments and airlines to address flight delay problems.

However, this study still has certain limitations: the research findings are based solely on 2023 flight data, and the limited duration of the dataset may affect the comprehensive understanding of long-term trends and seasonal changes. To conduct a more in-depth analysis of airport operation cluster patterns, future research should collect data with a longer time dimension to better capture the impacts of seasonal and annual changes. Meanwhile, consideration can be given to integrating flight data with other relevant data sources such as weather data and airline business data to further enrich the research dimensions.

\bibliographystyle{unsrt} 
%\bibliography{references} %%% Remove comment to use the external .bib file (using bibtex).
%%% and comment out the ``thebibliography'' section.

% Loading bibliography database
\bibliography{RefATMAIS}

%%% Comment out this section when you \bibliography{references} is enabled.

\end{document}